\documentclass[final,epsfig,amsfont]{article}
\usepackage{latexsym}
\usepackage{amsmath}
\usepackage{graphicx}
\usepackage{amssymb}
\DeclareGraphicsExtensions{.ps, .eps, .pdf, .png}
\newcommand{\eeq}{\end{equation}}
\newcommand{\beq}{\begin{equation}}
\newcommand{\nuq}[1]{\label{#1} \eeq}

\newtheorem{example}{Example}
\begin{document}
\title{Extreme value theory of evolving phenomena in complex dynamical systems:
firing cascades in a model of neural network}
\author{
Theophile Caby$^\dag$ \footnote{Aix Marseille Universit\'e, Universit\'e de Toulon, CNRS, CPT, 13009 Marseille, France} $\;$, Giorgio Mantica\footnote{Center for nonlinear and complex systems, dipartimento di scienza ed alta tecnologia, Universit\`a degli Studi dell' Insubria, Como, Italy.}
\footnote{INFN sezione di Milano, Italy.}
\footnote{Indam, Gruppo Nazionale di Fisica Matematica, Italy}}
\date{}
\maketitle
\begin{abstract}
We extend the scope of the dynamical theory of extreme values to cover phenomena that do not happen instantaneously, but evolve over a finite, albeit unknown at the onset, time interval. We consider complex dynamical systems, composed of many individual subsystems linked by a network of interactions. As a specific example of the general theory, a model of neural network, introduced to describe the electrical activity of the cerebral cortex, is analyzed in detail: on the basis of this analysis we propose a novel definition of neuronal cascade, a physiological phenomenon of primary importance. We derive extreme value laws for the statistics of these cascades, both from the point of view of exceedances (that satisfy critical scaling theory) and of block maxima. \\
{\em keywords: dynamical extreme value theory, complex systems, neural networks, criticality, neuronal avalanches}\\

\end{abstract}

\maketitle

\section{Introduction}
\label{sec1}

The dynamical theory of extreme values, well described in a series of works (see {\em e.g.} the review book \cite{lucabook} and references therein), has so far mainly considered low dimensional systems in a very simple setting that we will describe momentarily. To the contrary, its statistical ancestor, the so--called extreme value theory (EVT) \cite{gnede}, has been fruitfully applied to phenomena like earthquakes, floods, epidemics \cite{coles,epidemics}, that all involve complex, many dimensional interactions that operate for a certain period of time and are otherwise silent. It is the purpose of this paper to develop a theoretical framework to extend the dynamical treatment of extreme events to this kind of systems. We shall do this by focusing on a specific example, a model of neural network that has been introduced to describe various phenomena occurring in the electrical activity of the cerebral cortex \cite{shilni,rulkov,ruedi1,ruedi2}.

This model draws its origin, object and motivation from a vast body of existing research involving mathematics, physics and physiology. Nonetheless, the goal of this work is not to reveal new phenomena in the last field---although we will propose a new conceptual approach to a much studied problem, the statistical description of firing cascades, better known as neuronal avalanches \cite{plenz03,plenz04,plenz,beggs12,mazzoni}. Rather, the model is particularly suited to introduce a formal advancement in the dynamical theory of extreme values, which is also relevant to the neuronal case, among many others. In fact, at difference with standard theory, we will consider extreme events that occur in complex, many dimensional systems and that are not restricted to a specific instant in time, but which evolve during a finite time interval.

\subsection{Dynamical Extreme Value Theory}
The dynamical theory of extreme values can be summarized by considering a phase space $Z$, in which motions are determined by the repeated action of a deterministic map $\varphi:Z \rightarrow Z$ (that can also be probabilistic, but we will restrict ourselves to the first case). In addition, an {\em intensity function} $H : Z \rightarrow \mathbb{R}$ gauges the {\em value}, {\em i.e.} the intensity $H(z)$ of a phenomenon, an {\em event}, which occurs at the precise instant of time when the motion visits the point $z \in Z$. Extreme value theory is the study of the statistics of particularly intense events, {\em i.e.} large values of $H$, accumulated in time along trajectories of the motion. Without loss of generality we shall assume that time is discrete and denote it by the letter $n$. Trajectories of the system will naturally be written as $z_n = \varphi^n(z_0)$ and $I(n,z_0)$ will denote the intensity of the event occurring at time $n$ after the motion started at the point $z_0$: $I(n,z_0) = H(\varphi^n(z_0))$.

Originally \cite{lucabook}, very simple intensity functions $H$ have been investigated, which depend only on the Euclidean distance of $z$ from a single point $z^*$, and increase as this distance diminish. In this simplified setting extreme events are close approaches of the motion to such location. This situation was widely generalized by one of the present authors in \cite{gio}, by introducing intensity functions that depend on the distance from an {\em uncountable set of points} populating a fractal set $K \subset Z$. This analysis has recently been confirmed in full mathematical rigor \cite{jorge0}.

In addition, recall that phase space $Z$ and map $\varphi$ are complemented by an invariant measure $\mu$ to fully define a dynamical system. The measure $\mu$ determines the frequency by which the dynamics visit different regions of phase space. The combined action of this measure and of the geometry of $K$ in determining extreme value statistics was also investigated in \cite{gio}, revealing the r\^ole of fractal quantities like Minkowski dimension and Minkowski content, which were extended to the case of general ({\em i.e.} not necessarily Lebesgue and possibly singular) invariant measures.

Suppose now that the intensity of a phenomenon can{\em not} be determined {\em instantly} at time $n$, when the dynamics visits the phase--space point $z_n$, because this phenomenon develops over a certain time span, so that its true intensity can only be assessed when it terminates. This is the case of neuronal avalanches, when the electrical activity of the brain is sustained for a few milliseconds. But it is perhaps easier, to focus ideas, to consider a geophysical analogy \cite{hop}, to which the theory exposed herein can be equally applied.

An earthquake can be triggered by a single seismic event, a rupture, which propagates and multiplies in space and time. The total energy released by the quake is the intensity of the event, which can be assessed only after the {\em shock} terminates.
Notice at this point that even if this event takes place continuously over a finite time--span, it is  usually ascribed in geophysical catalogs to the moment of the first, triggering rupture. To put this observation into the theoretical framework of dynamical extreme values, the intensity function must take the form $H: Z^m \rightarrow \mathbb{R}$, in the sense that the event at time $n$ has an intensity $I(n,z_n)$ of the form $I(n,z_n) = H(z_n,z_{n+1},\ldots,z_{n+m-1})$, where $m$ is the duration of the phenomenon. This generalization of the conventional theory will be studied in this paper.

A first observation must be made at this point. Since we are in the presence of a deterministic dynamics, $z_{n+j} = \varphi^j(z_n)$, the intensity function takes on the special form

\begin{equation}
 H(z_n,z_{n+1},\ldots,z_{n+m}) = H(z_n,\varphi^1(z_n),\ldots,\varphi^{m-1}(z_n)) = {\cal H}(z_n).
 \label{eq-intens1}
\end{equation}

Intensity can therefore be thought of as depending only on the initial point $z_n$ via a global function ${\cal H}:Z \rightarrow \mathbb{R}$. This might lead one to think that the proposed approach is void of any generalization. As a matter of facts, because of the dependence of $H$ on future points of the evolution, the function $\cal H$ is presumably very complex and its {\em phase--space portrait}, {\em i.e.} its level sets, might feature a fractal, hierarchical organization, thereby providing a concrete dynamical example of the theory of \cite{gio} mentioned before. Moreover, when the dynamical system is chaotic, the function $\cal H$ in formula (\ref{eq-intens1}) cannot be given a more efficient computational scheme other than what expressed by the central expression in Eq. (\ref{eq-intens1}): the best one can do is to follow the dynamics as it evolves and record its data. Think again of the earthquake example: while it develops, it is {\em practically impossible} to forecast its intensity. On a more theoretical vein, recall the aphorism by Joe Ford: {\em a chaotic system is its own fastest computer and most concise description} \cite{joe}.

There is a further consideration that renders the generalization (\ref{eq-intens1}) far from trivial: the time--span of the phenomenon, $m$, is not known {\em a priori}, nor it is constant over different events: the integer $m$ in equation (\ref{eq-intens1}) {\em depends} on $z_n$. Because of the reasons just outlined there is no significantly better way to compute $m$ than to follow the dynamics until $m$ reveals itself: that is, the phenomenon under investigation---the quake---terminates.

\subsection{Neuronal avalanches}
It is now possible to return from the geophysical analogy to the paradigmatic example described in this paper: {\em neuronal avalanches}. It is observed that electrical stimuli originate (in the presence of, or without a sensorial stimulus) and propagate in the cerebral cortex to involve the {\em firing} of thousands of neurons. In fact, a neuron may either be silent, or {\em fire}, which means, polarize and send an electric stimulus via synaptic connections to other neurons. There is evidence that this activity composes a fundamental part of the computation performed by the brain \cite{halde05,shew,plenzi}. It can be recorded by suitable devices \cite{plenz} and it appears as bursts of firings, at times well separated between each other, but more often adjacent. In certain instances this activity may be due to synchronized oscillations of a large number of neurons \cite{hop0}, but more complicated patterns are frequently observed, somehow intermediate between the former and homogeneous, uncorrelated firing of individual neurons. A substantial amount of literature (see {\em e.g.} \cite{plenz03,plenz04,plenz,beggs12,mazzoni}) has shown that this regime can be described as {\em critical}, characterized by power laws and scaling relations. We focus our attention on both the critical and the subcritical regime, but we introduce a seemingly new operational definition of avalanche.

This definition is akin to what Rangan and Young \cite{lsy2,lsy3} have termed {\em multiple firing events}, as {\em the maximal sequence of firings that are caused by a single neuron}. By simulating each neuron in a network (typically, several thousands of these are considered) as characterized by microscopic variables (voltages and conductances) evolving according to coupled ODEs, 
and by the technical expedient of setting certain synaptic time constants to zero, Rangan and Young have been capable of determining exactly which neuron causes the firing (also called the {\em spike}) of a post-synaptic neuron.

The situation in the model that we will study is different and more general, in that the combined electrical activity of many different pre--synaptic neurons can cause the spike of a neuron.
Nonetheless, by studying the detailed dynamics of a single, unperturbed neuron---in particular, the {\em basin of attraction} of the spiking event---we will be able to introduce a {\em rule} according to which the firing of a particular neuron can be identified as the {\em precipitating cause} of the spike of another. 
We shall then define a {\em firing cascade} as the {\em tree} that mathematically encodes such chain of excitations.
We believe that this approach
is general enough to be applicable to a wide set of complex systems, composed of many interacting components.

Typical characteristics of these trees/cascades, like their intensity (number of nodes/number of firings), duration, and degree (number of levels in the tree) provide intensity functions $H$ to which extreme value theory can be applied, both from the point of view of exceedances and of block maxima (concepts to be made precise below), thereby providing a concrete example of the abstract theory described in the previous section.

This analysis will produce scaling relations that also permit to compute the probability that all cascades in a certain time interval have intensity less than a given threshold. This kind of extreme value laws for the size of neuronal cascades has not been investigated before, to the best of our knowledge---see however \cite{leili} for a discussion of the relevance of extreme value theory to sensorial perception.

\subsection{Contents of the paper}
The plan of this paper is the following. We first describe in Sect. \ref{sec2} the dynamical model of neural network that we adopt, borrowed from \cite{ruedi2}. It consists of a collection of two--dimensional maps, each representing an individual neuron, initially proposed by Rulkov \cite{shilni,rulkov}. In sections \ref{ssec-unp} and \ref{sec-netw} the theory follows closely, with minor modifications, the original references \cite{shilni,rulkov,ruedi1,ruedi2}. This is a necessary introduction, not only to set notation and define concepts, but also to investigate a few details of dynamical import: in particular, the {\em basin of attraction of the spiking event}, described in Section \ref{ssec-pert}, which is instrumental to define causality interactions.
Neurons do not work in isolation, but are linked in a sparse network, where different neurons interact only when the spiking state is reached. In Sect. \ref{sec-netw} we describe the model of network connections,  following \cite{ruedi1,ruedi2}, and then in Sect. \ref{sec-netdyn} the collective dynamics that takes place on it.

We start in Sect. \ref{ssec-invm} from the problem of the existence of an invariant measure that we heuristically replace by Birkhoff time averages, having nonetheless in mind that this procedure might be dubious due to possible long transients in the dynamics \cite{politi1}.  Section \ref{sec-aval} contains the fundamental definition of {\em firing cascades as trees of causal excitations}, as well as its practical implementation, which is based on the concept of basin of attraction of the spiking event. This permits to define and measure important quantities like the intensity $S$ of a cascade, its degree/generation level $G$ and its time span $T$.  In Sect. \ref{sec-quantities} the statistics of these quantities is analyzed, showing a transition to critical behavior as the intensity of coupling among different neurons is increased; the exponents of the power--laws describing the above quantities obey critical scaling theory and agree with previous experimental findings and theoretical research. From the point of view of extreme value theory, this analysis can be classified as the study of exceedances.

Finally, in Sect. \ref{sec-evt} we further widen the scope of our investigation: we show the emergence of Extreme Value Laws in the statistics of extrema in time--intervals of finite length. That is, we will derive scaling relations for the distribution of the {\em largest} event in a certain time interval, both in the critical and the sub--critical regime. Numerical experiment will confirm the formal theory. In the considered example, this offers a new statistical characterization of neuronal cascades, which might perhaps be of physiological relevance \cite{leili}.

The conclusions briefly summarize the work, and an Appendix describes in an elementary way the stability regions, in parameter space, of the fixed point of Rulkov map.

\section{The dynamical model: a sparse network of Rulkov maps}
\label{sec2}

A variety of models have been proposed to study numerically neuron dynamics. Many of these are random. Purely dynamical models, on the other hand, can be continuous time models, involving systems of differential equations \cite{hod}. They are able to reproduce a whole variety of observed biological patterns \cite{lsy2,lsy3,lsy}, but are expensive from a computational point of view. On the contrary, Rulkov 2d--map \cite{shilni,rulkov} is a good compromise between capacity of well reproducing real biological behaviors, such as irregular spiking and bursts of spikes \cite{rulkov,ruedi1} and fast computations, two features that are crucial for analysis.
In this section, we briefly describe the dynamics of a single neuron, which is effected by Rulkov map. We follow the implementation presented in  \cite{ruedi1,ruedi2} that also contains illustrative pictures. We need only to mention that, at difference with these references,
we do {\em not} allow a random external input: our dynamical system is fully deterministic.

Rulkov map describes the evolution of a fast variable $x_n$ ($n$ is integer time) that models the membrane potential of a neuron, and a slow control variable $y_n$ that, although it is void of physiological meaning, is required to pilot the evolution of the former, so to reproduce some patterns observed in real experiments. This pair of variables describes the state of every neuron in the system. We label neurons by the variable $i=1,\ldots,N$, so that the phase space of the system is $Z = \{(x^i,y^i), i=1,\ldots,N\}$. Recall that $n$ labels time;
Rulkov map is then defined as follows:
\beq
\left \{
\begin{array}{ll}
    x^i_{n+1} $ = $  F(x^i_n,y^i_n,I^i_n,x^i_{n-1}) \\
    y^i_{n+1} $ = $ y^i_n - \mu (x^i_n+1) + \mu \sigma + \mu I^i_n.
\end{array}
\right .
\nuq{eq-map1}
The variable $I^i_n$ is the synaptic input, a {\em current}, acting on neuron $i$ at time $n$, defined below in Eq.  (\ref{eq-fir1}). Different neurons are coupled only via this term.
It is convenient to first discuss the dynamics when this external input is set to a constant (that may be taken to be zero, for sake of definiteness)  and each neuron evolves independently of the others.
\subsection{Unperturbed dynamics of the single neuron}
\label{ssec-unp}
In this subsection, accounting for the fact that neurons do not interact, we drop the neuron index $i$ for clarity. The function $F$ in Eq. (\ref{eq-map1}) is defined as follows:

\begin{equation}
F(x_n,y_n,I_n,x_{n-1}) = \left \{
\begin{array}{ll}
    \frac{\alpha}{1 -x_n} + y_n+\beta I_n      & $ if $  x_n \leq  0 \\
    \alpha+y_n+\beta I_n & $ if $   x_n > 0 $ and $  \{x_n < \alpha+y_n+\beta I_n $ and $  x_{n-1} \leq 0 \}\\
    -1 & $ if $ x_n > 0 $ and $ \{ x_n \geq \alpha+y_n+\beta I_n $ or $ x_{n-1} > 0 \}
\end{array}
\right .
 \label{eq-map1b}
\end{equation}

The parameters $\alpha,\beta,\mu,\sigma$ appearing in equations (\ref{eq-map1}),(\ref{eq-map1b}) are all positive. 

The parameter $\beta$ modulates the relative influence of the synaptic input $I_n$ with respect to $y_n$ in Eq. (\ref{eq-map1b}). We choose $\beta=0.133$.
Notice that a memory effect is present in the above definition, via the value of the variable $x$ at time $n-1$.

\begin{figure}
  \begin{center}
    \begin{tabular}{cc}
      \resizebox{60mm}{!}{\includegraphics[height=60mm, angle=-90]{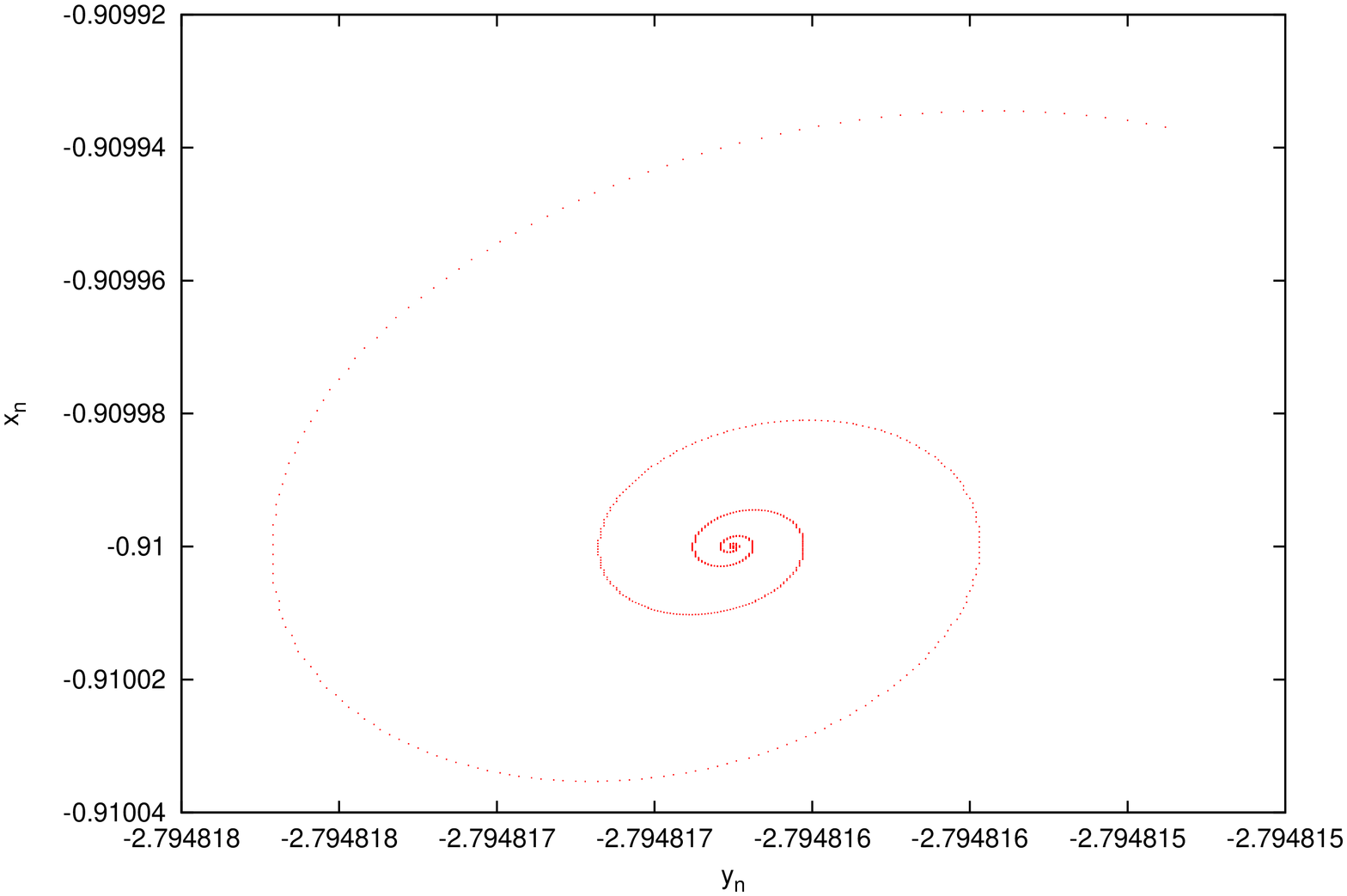}}   &
 \resizebox{60mm}{!}{\includegraphics[height=60mm, angle=-90]{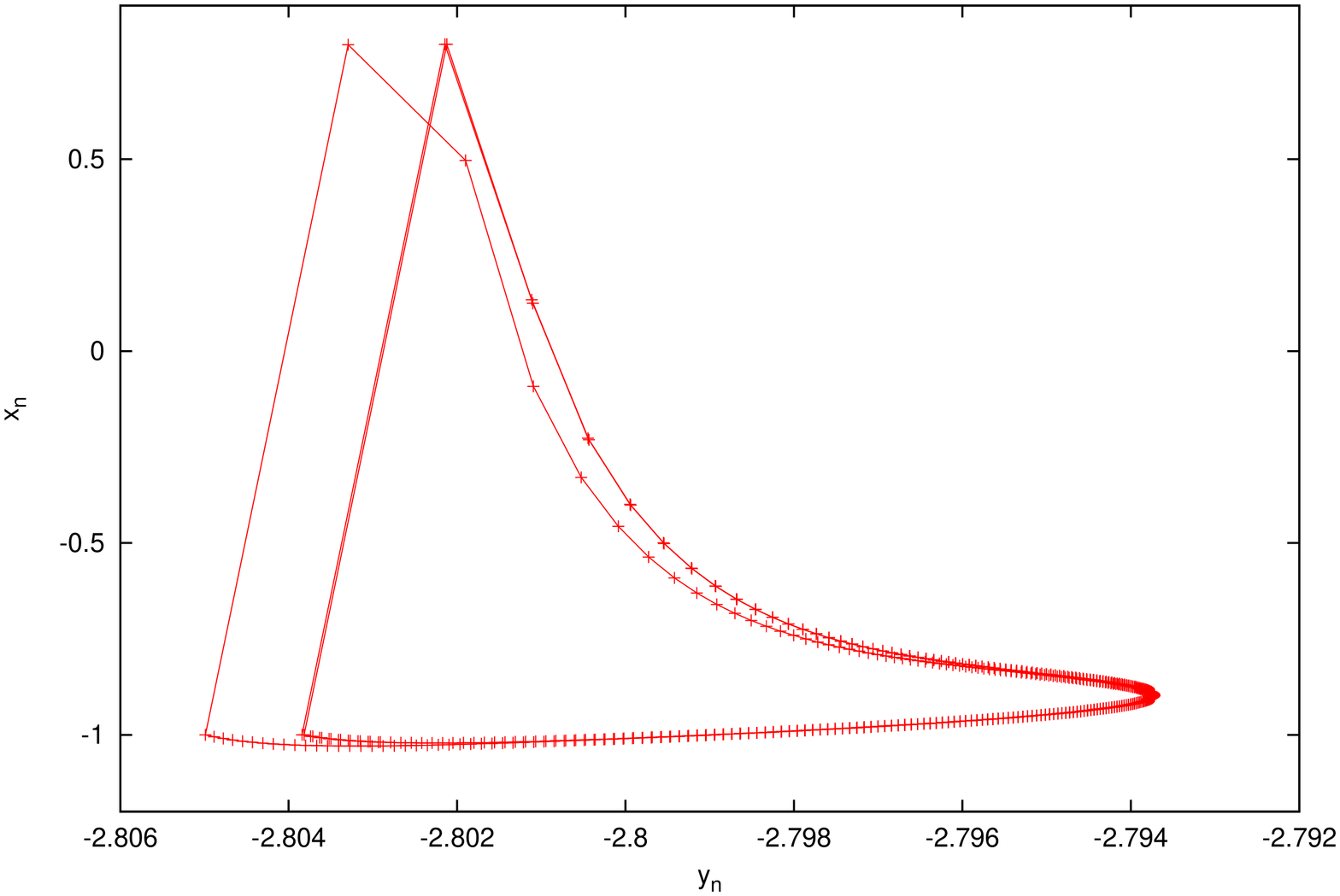}}  \\
    \end{tabular}
    \caption{Typical trajectories in the single--neuron phase space $(y,x)$ of an isolated INSN (left) and of an isolated ISN (right). Lines between points are merely to guide the eye. Motions happen counter-clockwise in both frames. Notice the different range of the axes; in the INSN case the figure is zoomed about the fixed point of the map.}
    \label{fig-traj1}
  \end{center}
\end{figure}

The parameter $\mu$ is taken to be small ($\mu = 10^{-3}$), so that the variable $y$ evolves slowly with respect to time, following the second equation in (\ref{eq-map1}). Also, its values cover a limited range.
The combination of $\mu$ with $\alpha$ and $\sigma$ determines the structure of motions, as discussed in detail in \cite{shilni,rulkov}. In Appendix we sketch a simple analysis of the stability of the fixed point $(\sigma-1,\sigma-1+\frac{\alpha}{\sigma-2})$. Stoop {\em et al.} \cite{ruedi2} choose $\alpha=3.6$, so that only $\sigma$ is left free to vary. As shown in Appendix, under these conditions the map admits a bifurcation at $\sigma=\sigma_{cr}=2-\sqrt{\frac{\alpha}{1-\mu}}$. Parameter space is accordingly parted in four regions, I to IV, plotted in Fig. \ref{fig-regio} in the Appendix.
Choosing $\sigma=0.09$ puts the parameter-space point in region II, implying an attracting fixed point, while $\sigma=0.103$ leads to region III, where the fixed point is unstable.
This permits to define two kinds of neurons: Intrinsically Non-Spiking Neurons (INSN) (parameters in region II, motion in the left frame in Fig. \ref{fig-traj1}) and Intrinsically Spiking Neurons (ISN) (region III, right frame). These latter are the initiators of the dynamics of the network. The different behaviors are apparent in Fig. \ref{fig-traj1} where we plot motions in the single--neuron phase space (as in Fig. 1 in \cite{ruedi2}, the $y$ axis is in abscissa and the $x$ axis in ordinate).

A neuron is said to {\em fire} when its membrane potential $x_n$ reaches a local positive maximum (depolarization) and then suddenly drops to the hyperpolarized state $x=-1$: in physiology one talks of increasing polarization when the potential becomes more negative. At this moment a signal 
is transmitted to the other neurons, as described in the next two sections.
It is instructive to follow the motion in the right panel of Fig. \ref{fig-traj1}. Three {\em cycles} are plotted (two are almost coincident and require attention to be distinguished),  each of which contains a spike. Start from the rightmost part of a cycle and follow points rotating counter-clockwise, so that $y_n$ initially decreases and $x_n$ increases. This happens because of the first alternative in Eq. (\ref{eq-map1b}).  The map (\ref{eq-map1}) is defined in such a way that when $x$ becomes positive at time $n$ ($x_{n-1}$ being negative: draw a horizontal line at ordinate $x=0$ in the figure for visualization), either $x_n$ is larger than $\alpha+y_{n}+\beta I_{n}$, so that at the next step the membrane potential drops to the negative value $x_{n+1}= -1$ (third alternative), or else it further increases to the value $x_{n+1} = \alpha+y_{n}+\beta I_{n}$ (second alternative), before achieving the value $x_{n+2}=-1$ at the following iteration, again in force of the third alternative.
This motivates the introduction of the spike variable $\xi_n$ that is always null, except at those times when the neuron potential reaches the hyperpolarized state, in which case it takes the value one. It can be formally defined as follows (see eq. (\ref{eq-map1b})):
\beq
\xi_{n+1} = \left \{
\begin{array}{ll}
     1 $ if $ x_n > 0 $ and $ \{ x_n \geq \alpha+y_n+\beta I_n $ or $ x_{n-1} > 0 \} \\
    0 $ otherwise$.
\end{array}
\right .
\nuq{eq-spike1}
The spike variable $\xi_n$ will be used in Sect. \ref{sec-netw} to define the dynamical action of synaptic coupling.

\subsection{Perturbed dynamics of the single neuron: basin of attraction of the firing event}
\label{ssec-pert}
The synaptic input $I_n$ received by a neuron can either be stochastic or deterministic, as in our case. How this latter is generated in our model is inessential for the present discussion and will be explained in the next section;
suffices here to observe that $I_n$ enters equation (\ref{eq-map1b}) linearly. In particular, a positive current $I_n$ increases the potential $x_n$ (hence, it depolarizes the neuron) and can therefore be associated with an {\em excitatory} action, while the reverse is true for a negative current $I_n$. Notice however that this rule of thumb has non--trivial exceptions that we will describe momentarily.

\begin{figure}
  \begin{center}
    \begin{tabular}{cc} \resizebox{60mm}{!}
    {\includegraphics[height=60mm, angle=-90]{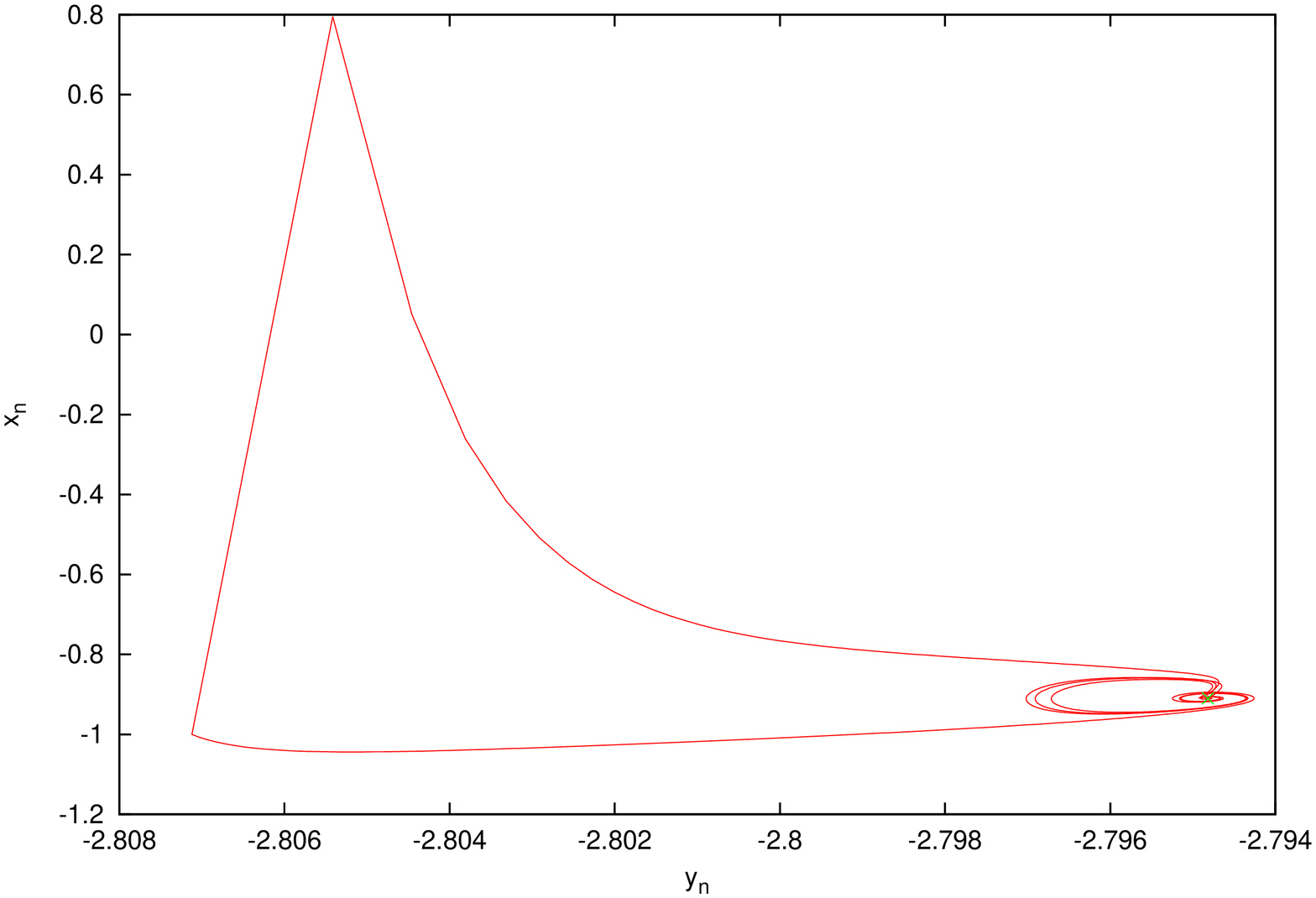}}&
      \resizebox{60mm}{!}
      {\includegraphics[height=60mm, angle=-90]{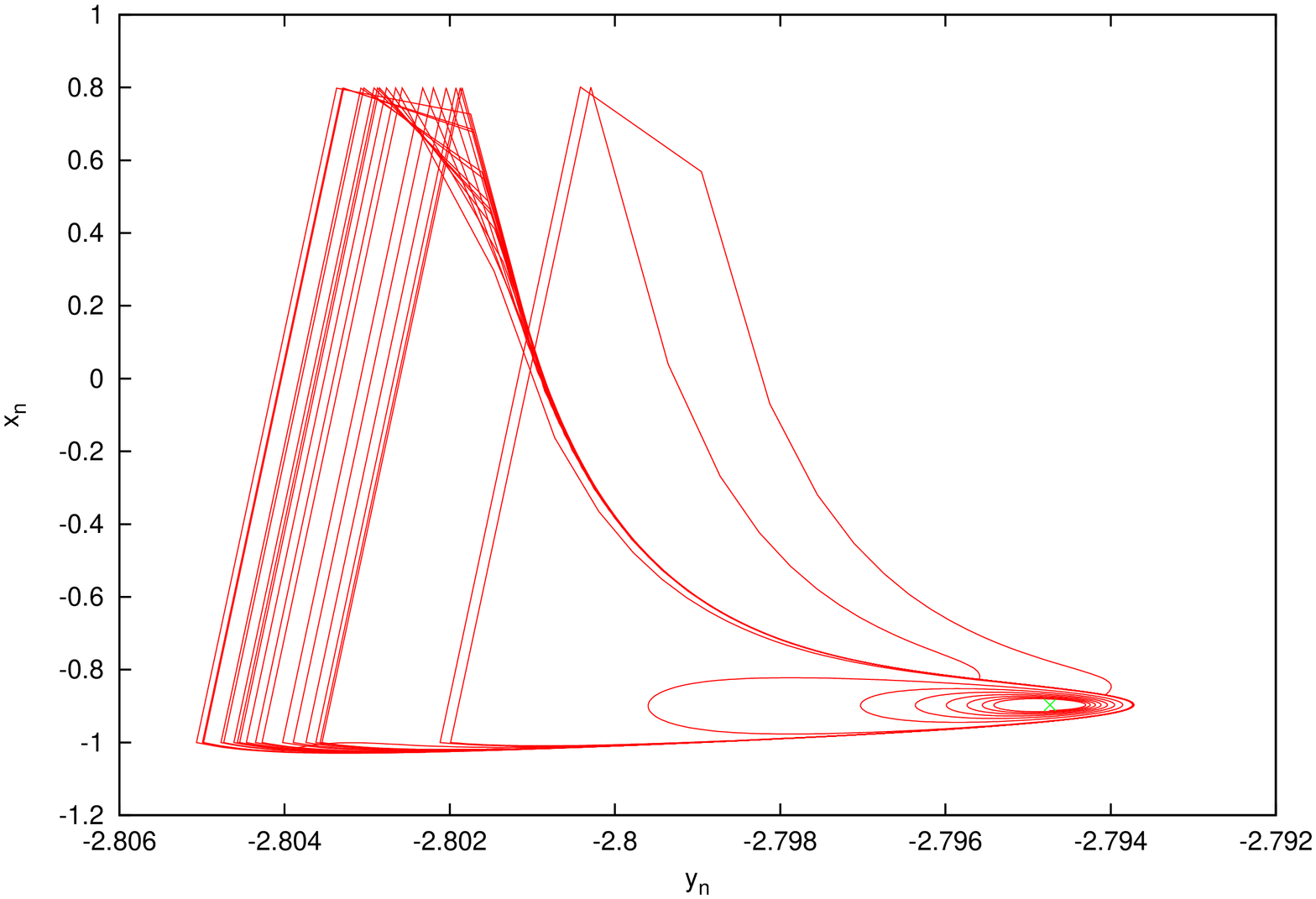}}
      \\
    \end{tabular}
    \caption{Typical trajectories in the single--neuron phase space $(y,x)$ of INSN (left) and ISN (right) neurons embedded in a network. Marked by a symbol are the fixed points of the Rulkov map. Motions takes place counter-clockwise in both frames.}
    \label{fig-trajINS}
  \end{center}
\end{figure}

\begin{figure}
  \begin{center}
    \begin{tabular}{cc} \resizebox{60mm}{!}{\includegraphics[height=60mm, angle=-90]{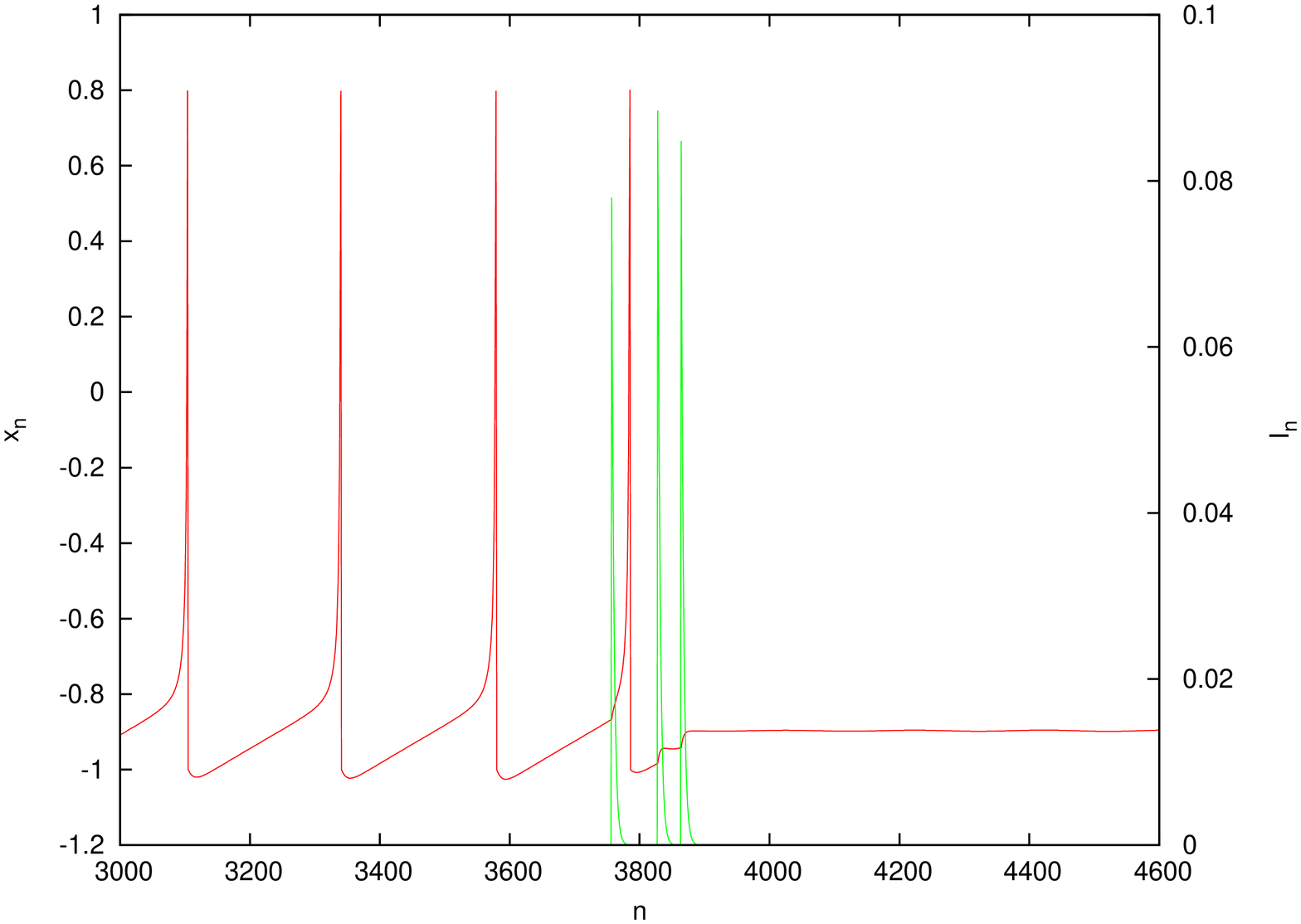}}&
      \resizebox{60mm}{!}{\includegraphics[height=60mm, angle=-90]{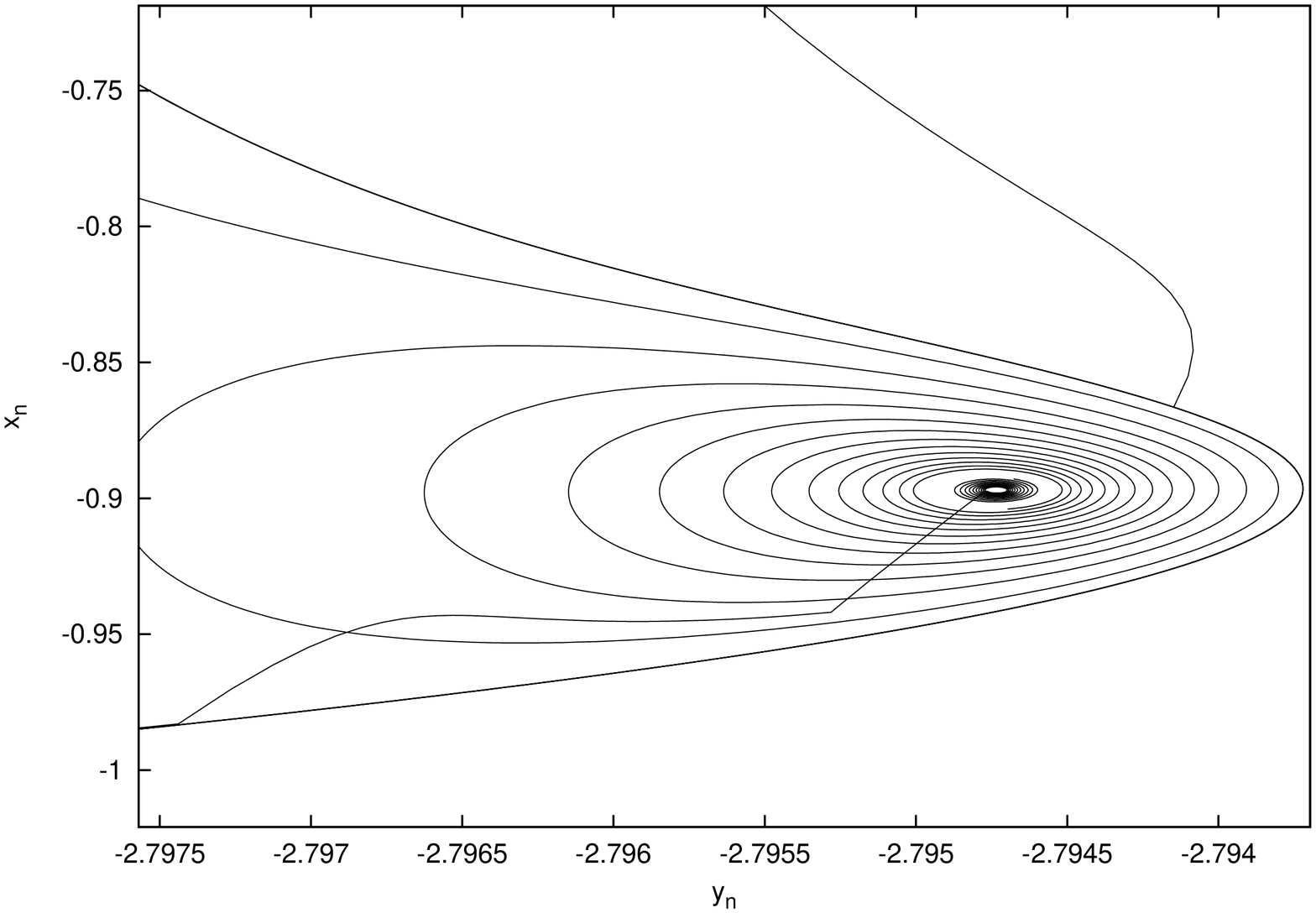}}
      \\
    \end{tabular}
    \caption{In the left frame, the potential $x_n$ of an ISN (red curve) is plotted versus time $n$, together with $I_n$ (green curve). The right frame plots the corresponding phase--space trajectory. See text for further discussion.}
    \label{fig-capt}
  \end{center}
\end{figure}

Figure \ref{fig-trajINS} shows how the dynamics of ISN and INSN observed in Fig.  \ref{fig-traj1} is perturbed by $I_n$. Typical trajectories are shown, in which sharp angles appear when $I_n$ is non--null. Notice however that time being discrete, lines in the figure join distinct points--albeit typically very close, so to simulate a continuous trajectory that does not exist. In the left frame one can observe how INSN, silent when isolated, can be triggered to spike by the input $I_n$. 
The dynamics of ISN (right panel) can also be perturbed, altering their regular spiking regime. In fact, consider Fig. \ref{fig-capt}, left panel, where the potential $x_n$ of an ISN (red curve) is plotted versus time $n$, together with the synaptic input $I_n$ it receives (green curve).
The synaptic current $I_n$ is positive, stemming from an excitatory action. Three excitations are visible in the range displayed. Prior to these the ISN spikes regularly with a period of about 242 iterations. The first excitation has the effect of anticipating the successive spike by 35 iterations. In the right panel, which shows the same motion in the single--neuron phase space, regular oscillations appear as a thick limit cycle. The first spike is evident in the thread which emerges from the cycle at the right of the picture, and moves initially by increasing both $x_n$ and $y_n$, according to the fact that the excitation $I_n$ is positive: see Eq. (\ref{eq-map1b}). Returning to the left panel, we notice two more excitatory inputs to the neuron that have the paradoxical effect of suppressing further spikes! This can be explained by detecting, in the right panel, two more sharp angles in the trajectory, the first starting form the bottom left of the picture. In both cases $x_n$ and $y_n$ increase, but this has the effect of pushing the trajectory towards the fixed point of the map, which is unstable but from whose neighborhood the motion requires a long time to escape.

We have so discovered that excitatory actions can sometimes produce an inhibition of spikes, due to the particular nature of the phase space of the system.
For a similar reason, inhibitory actions can sometimes induce spikes. This paradox leads to a deeper understanding of the dynamics of this system.

\begin{figure}
\centerline{\includegraphics[height=10cm]{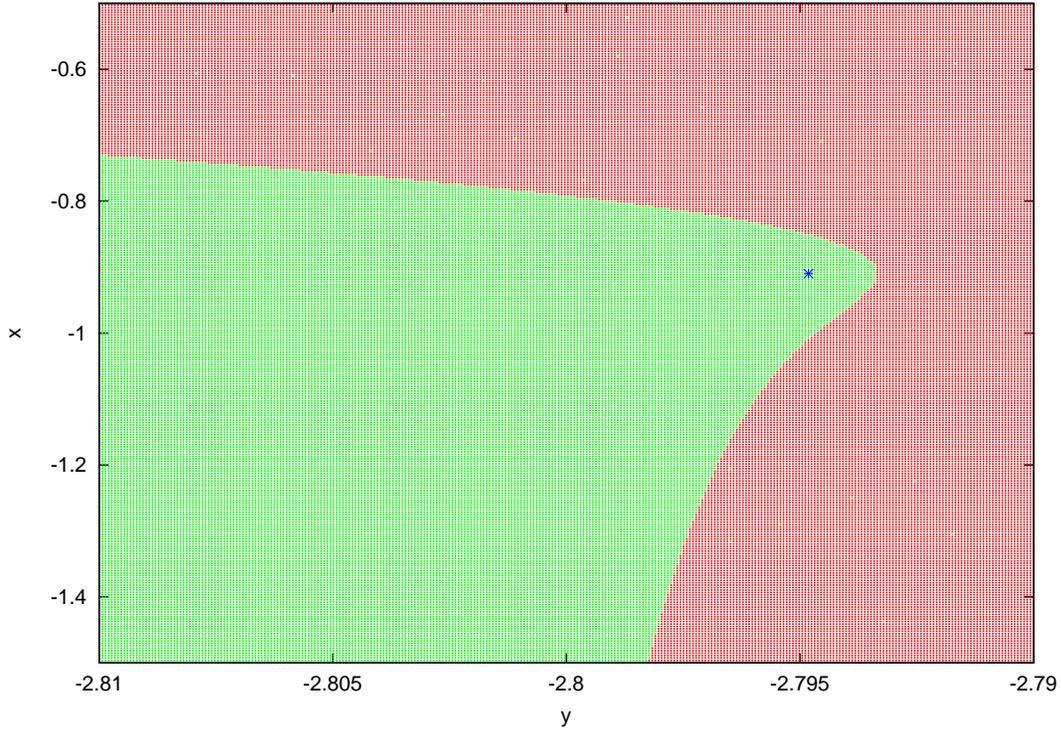}}
\caption{Basin of attraction of the spiking event, in red, and basin of attraction of the fixed point in green. The fixed point is marked in blue.}
\label{fig-basin}
\end{figure}

In fact, consider an INSN, like those pictured in the left panels of Figs. \ref{fig-traj1}, \ref{fig-trajINS}. Let $(y,x)$ be the initial condition of the motion and let $I_n =0$ (isolated neuron). Even if finally attracted by the fixed point, some trajectories will spike before that happens. In the left frame of Fig. \ref{fig-basin} we color in red such initial conditions, and in green those that head directly to the fixed point, whose position is marked by a blue asterisk. A limit cycle separates the two regions. In the right frame, the phase--space portrait near the stable fixed point is displayed.
It is then clear that an inhibitory action (reducing both $x_n$ and $y_n$) can move a point from the green to the red region and that an excitatory {\em kick} can do the opposite, but it is equally clear that this occurs only in a restricted region of phase space. In fact, increasing $x$ typically moves a phase--space point from the green to the red region, {\em i.e.} from the basin of the fixed point to that of the spike. In particular, when the potential $x$ of a NISN is larger than a certain threshold $x_{th}$, the paradoxical effects just described are not possible. In fact, all initial conditions with $x_n$ larger than $x_{th}$ lie in the red region and lead to firing, even in the absence of further excitation: the effect of the current $I_n$ that puts a NINS in the red region can therefore only be reversed by an inhibitory input.
This fact will be of paramount importance in the following.
From the analysis of Fig.  \ref{fig-capt} we deduce that, in the range of parameters of our experiments, we can safely take $x_{th} = -0.7$.

\begin{figure}
\centerline{\includegraphics[height=10cm,angle=270]{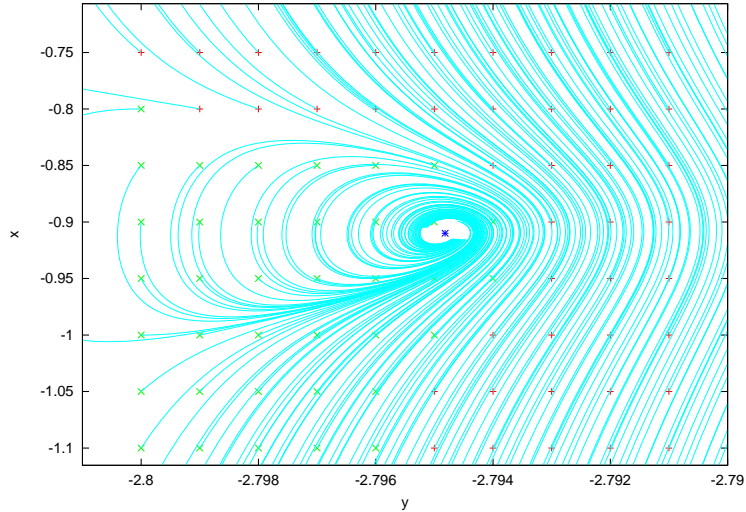}}
\caption{Detail of figure \ref{fig-basin} with trajectories plotted.}
\label{fig-basin2}
\end{figure}

\subsection{Network topology and synaptic coupling}
\label{sec-netw}

According to the way in which they act upon others, neurons can either be classified as {\em excitatory} or {\em inhibitory}. Following \cite{ruedi2}
we consider a network of size $N$ with a ratio of inhibitory/excitatory neurons $r$, so that the first $n_{ex} = \lfloor{Nr}\rfloor$ neurons are excitatory and the remainder $N - n_{ex} = n_{in}$ are inhibitory.  Furthermore, as described in the previous section, neurons can either be intrinsically spiking or not, so that each group (excitatory and inhibitory) can also be parted in two: out of the $n_{ex}$ ($n_{in}$) excitatory (inhibitory) neurons $m_{ex}$ ($m_{in}$) are chosen to be intrinsically spiking.

Neurons are connected so that each of them is acted upon by $p_{ex}$  excitatory neurons and $p_{in}$ inhibitory neurons, chosen at random in the network---with no regard on whether ISN or NISN, with the only constraint that no neuron can act upon itself.
The six integer numbers $n_{ex}$, $m_{ex}$, $n_{in}$, $m_{in}$, $p_{ex}$, $p_{in}$, plus the connection diagram fully specify the network. Analytically, this is formalized by the adjacency matrix $\omega_{ij}$, defined by

\begin{equation}
\omega_{ij} = \left \{
\begin{array}{ll}
   0  &\mbox{ if  neuron } j \mbox{ does not act on neuron } i, \\
    1  &\mbox{ if  neuron } j \mbox{ has excitatory action on neuron } i, \\
   \psi &  \mbox{ if  neuron } j \mbox{ has inhibitory action on neuron } i.
\end{array}
\right .
 \label{connect}
\end{equation}

The factor $\psi>0$ serves to differentiate the intensity of inhibitory with respect to excitatory action. We choose $\psi=3$.

Coupling in the network is driven by action potentials generated by firing neurons.
Following again \cite{ruedi2} the spike of a presynaptic neuron $j$ determines the input $I_n^i$ received by the neuron $i$ according to the linear difference equation:
\beq
I_{n+1}^i - \eta I_n^i =  W \sum_{j=1}^N \omega_{ij} (\chi_{ij}-x_n^i) \xi_n^j.
 \nuq{eq-fir1}
$W$ is a crucial intensity parameter, varying which we can increase the effect of synaptic excitation: recall that $I^i_n$ enters the dynamical equation of evolution of neuron $i$, Eqs. (\ref{eq-map1}),(\ref{eq-map1b}).
The form of Eq. (\ref{eq-fir1}) can be explained in physiological terms. First observe that the homogeneous equation (with null r.h.s.) implies an exponential decay: $I_n \sim \eta^n$ (we choose $\eta = 0.75$).
The spike variables $\xi_n^j$ at r.h.s. are the source of the currents $I^i_n$.  In Eq. (\ref{eq-fir1}) we notice that when the synapse reversal potential $\chi_{ij}$ is larger than the potential $x_n^i$, its activation (that happens when $\xi_n^j = 1$) increases the synaptic input $I_n^i$ and also, because of Eq. (\ref{eq-map1b}), the membrane potential $x^i_{n+1}$. This is equivalent to a depolarization of the neuron and therefore the synapse acts as excitatory. In the opposite case, the activation polarizes the neuron and the action is inhibitory. The numerical values of the above parameters are chosen as in \cite{ruedi1,ruedi2}, where their physiological relevance is discussed:
the reversal potential is null for excitatory connections ({\em i.e} $\chi_{ij} = 0$ when neuron $j$ is excitatory) and equal to $\chi_{ij} = -1.1$ when neuron $j$ is inhibitory, for all $i$.
This ends the description of the {\em laws of motion} of the network.

\section{Network dynamics}
\label{sec-netdyn}
We can now study the dynamical behavior of the full network. We start by discussing a basic problem: the existence of an invariant measure. Then, we turn to the definition of {\em firing cascades}.

\subsection{Invariant measure}
\label{ssec-invm}
Equations (\ref{eq-map1}),(\ref{eq-map1b}),(\ref{eq-spike1}),(\ref{connect}) and  (\ref{eq-fir1}) define a multi-dimensional dynamical system in the phase space $Z = \mathbb{R}^{2N}$ with variables $(x^i,y^i), i =1,\ldots,N$. Let $z=(x,y)$ be such $2N$--component phase--space vector and let $\varphi$ be the dynamical evolution operator. A fundamental ingredient of dynamical theory are the so--called invariant measures, that is, positive Borel measures that are preserved by the dynamical evolution. Lacking a theoretical control of these measures in many--dimensional systems like the one under consideration, we substitute phase--space averages by Birkhoff sums over trajectories. In measure--preserving systems, the two coincide almost everywhere (with respect to the initial point of a trajectory, under the invariant measure).

More precisely, if we consider an intensity function $H(z)$ (for simplicity of notation, from now on we write $H$ for ${\cal H}$, see Eq. (\ref{eq-intens1})), we construct its distribution function $F_H(h)$ as follows. We start from a random point $z^* \in Z$. We iterate the dynamics $\varphi$ for a number of iterations $n_0$ sufficiently large to let the system reach a sort of dynamical equilibrium.
At that moment we let $z_0 = \varphi^{n_0}(z^*)$ be the initial point of a trajectory and we compute the sample values $\{ H(\varphi^{n} (z_0)), n=0,\ldots,L-1\}$, by which the distribution function $F_H$ is
\beq
     F_H(h) = \lim_{L \rightarrow \infty}
     \frac{1}{L} \# \{ H(\varphi^{n} (z_0)) \leq h, \; 0 \leq n < L \}.
\nuq{eq-disth}
Clearly, in numerical experiments the limit is replaced by the value obtained for a large but finite sample length $L$. We shall also use the complementary distribution function $E_H(h) = 1 - F_H(h)$: this is the measure of the tail of the distribution of values of $H$:
\beq
     E_H(h) = \mu (\{ z \in Z \mbox{ s.t. } H(z) > h \})
\nuq{eq-disth2}

The study of coupled map lattices has revealed puzzling phenomena like stable motion with exponentially long chaotic transients \cite{politi1,politi2}, so that the numerical procedure just explained might be theoretically flawed. Nonetheless, the result of \cite{ruedi2} on small networks ($N=128$) revealing a transition to positive Lyapunov exponents lead us to believe that convergence in Eq. (\ref{eq-disth}) can be readily obtained.

\subsection{Firing cascades as trees of causal excitations}
\label{sec-aval}

A crucial difference between conventional coupled map lattices and the network under study is that in this latter interaction between different maps happens only at specific instants of time, while for most of the time individual systems evolve independently of each other. When it comes to physiological networks, the first and foremost consequence of this fact is that information can be stored, elaborated and transmitted \cite{plenzi}. In a dynamical perspective this calls attention to collective behaviors like synchronization \cite{hop0,arka}. Here we focus on a more elusive set of phenomena: neuronal avalanches, which we prefer to call firing cascades.

\begin{figure}
\centerline{\includegraphics[height=10cm, angle=270]{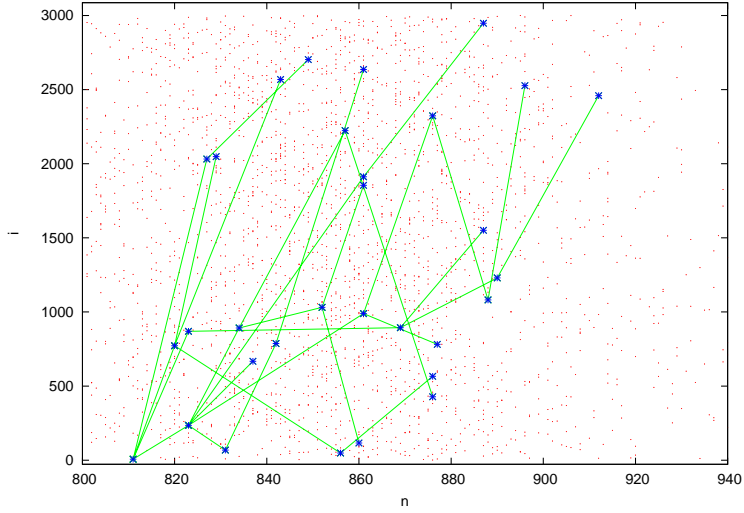}}
\caption{Tree construction. Red dots mark neuron firing events $(n,i)$; green lines join causally related events (blue symbols). See text for description.}
\label{fig-tree}
\end{figure}

Most of the studies on the collective behavior of neural networks define avalanches as the uninterrupted activity of the network during a certain time interval \cite{plenz03,mazzoni}. Statistical properties of these collective events have been extensively studied in a large series of works, both experimental, see {\em e.g.} \cite{plenz03,plenz04,plenz,beggs12,mazzoni}, and theoretical, with the aid of model systems \cite{lsy2,lsy3,lsy,lucilla1,shew,ott12,eu02,theo07,halde05,lucilla2}, but statistical analysis alone does not clarify the dynamical detail of the activity network and, in particular, of the impulse propagation. Moreover, it suffers from a degree of arbitrariness in the definition of avalanche. In fact, coarse--graining of time is usually employed, for spiking events appear at particular instants of time and must be grouped to define the quiescence periods separating different avalanches. Figure \ref{fig-tree} shows a conventional {\em raster plot} of spikes, depicted as red points at coordinates $(n,i)$ where $n$ is the time at which neuron $i$ spikes. One observes regions with different density of spikes. The usual procedure \cite{plenz03} is to partition the time axis into different bins, so that a sequence of adjacent, non--empty bins constitutes a cascade. The arbitrariness in the choice of the bin length is apparent, and its influence on the resulting statistical properties has been well described in \cite{plenz03}. Moreover, different regions of the network even if uncorrelated but spiking at close instants may be classified in the same cascade.
In short, the true dynamics of the network is overlooked in this approach. This is justified in the analysis of experimental data, when no access to the network of connections is available.

We pursue here a different approach, based upon the detailed knowledge of the dynamics presented in the previous sections.
We have remarked that different types of neuron are present in the network: in particular, excitatory ISN are the origin of propagating excitations. When one of these neuron spikes, say neuron $j$, it acts upon its post-synaptic neurons via Eq. (\ref{eq-fir1}). As observed in Section \ref{sec2}, if such postsynaptic neuron is also an ISN, this action will alter its dynamics, anticipating or delaying its next spike. Otherwise, if the post-synaptic  neuron $i$ is of the kind NISN, this excitation might eventually drive it to spike, this occurrence following from the dynamics generated by Eq. (\ref{eq-map1}). The induced spiking, when it takes place, is not instantaneous:  according to the coordinates $(x_n^i,y_n^i)$ at the time of excitation, the motion requires a certain amount of time to exit the region close to the fixed point and spike. It is nonetheless possible to predict if this event will take place, using the notion of  basin of attraction of the spiking event 
discussed in Sect. \ref{sec2}. Recall that this is defined as the set of initial conditions of the {\em isolated non--spiking} neuron which lead to a spike, before returning to approach the fixed point. When $x_n$ is pushed by a synaptic excitations into this basin (red in Fig. \ref{fig-basin}) the NISN will fire, regardless of further excitatory actions: such spike can only be avoided by inhibitory actions.

This reasoning permits to draw a {\em causal} relationship between the spiking of a NISN and {\em a single originating excitation} from another neuron. 
Even if it is clear that in a complex network many pre--synaptic neurons can contribute to set the dynamics of $i$ on course to firing, we {\em stipulate to ascribe the cause of the spike of the non--intrinsically spiking neuron $i$ to a single excitatory neuron, $i'$, the one which is the last to act on $i$, at time $n'$, before it enters the basin of attraction of the spiking event, to eventually fire at time $n$}.

Excitations following this time are not considered in this count, because they are not necessary to induce a spike, which would happen nonetheless. Equally not considered are impulses preceding such precipitating event---the straw that broke the camel's back---because if taken alone they might not be enough to cause firing. As a simpler rule, that can also be applied in experiments, we approximate the basin of attraction by the fact that $x^i$ exceeds a threshold value $x_{th}$. This procedure can be formulated as follows: we say that {\em the spike of the excitatory neuron $i'$ at time $n'$ causes the spike of the INSN $i$ at time $n$, and we write $(n',i') \rightarrow (n,i)$ if}

\beq
\begin{array}{ll}
 (n',i') \rightarrow (n,i)  \Longleftrightarrow & \xi_{n'}^{i'} = 1, \;  \xi_n^i = 1,\; \\
  & \exists \; n^*, \; n' < n^* < n, \mbox{ such that the following holds:} \\
  &
  x_m^{i} < x_{th} \mbox{ for } n' \leq m < n^*, \; \\
 &  x_m^{i} \geq x_{th} \mbox{ for }   n^* \leq m < n,  \\
   & \xi_m^j = 0 \mbox{ for }  n' < m < n^*, \; \forall j  \mbox{ such that }  \omega_{i,j} = 1.
   \end{array}
\nuq{eq-chain1}

The event that two neurons spike simultaneously and excite the {\em same} postsynaptic neuron is fully negligible and can be disposed by an {\em ad--hoc} prescription.
Following rule (\ref{eq-chain1}) we can therefore create trees of excitations, which start at the firing of a single excitatory ISN and propagate to INSN's through the network.  Any such tree is defined to be a {\em firing cascade}. Return for illustration to Fig. \ref{fig-tree}, which also shows a realization of this construction. Green lines $(n',i') \rightarrow (n,i)$ join causally related spikes  (blue symbols), on the background of red dots that mark all spiking events.
The picture shows how the initial firing of an excitatory, intrinsically spiking neuron ($i'=7$ at $n'=811$) lead non--intrinsically spiking neurons to fire ($i=772$ at $n=820$, $i=235$ and $i=869$ at $n=823$, $i=2032$ at $n=827$) that, on their turn, stimulate additional NISN's  (for a total of 31 spikes in the figure). Clearly, not all {\em physical} connections are {\em dynamically} activated following a spike, for otherwise the cascade will have no end. In the example, neuron $i'=7$ is also presynaptic to $i=908$, $i=1115$ and $i=2677$ that remain silent.

\section{Statistical analysis of firing cascades}
\label{sec-quantities}

The dynamical approach described in the previous section permits to define precisely and measure accurately various quantities usually defined via time binning. These are:
\begin{itemize}
\item $S$, the size of a cascade, that is, the total number of spikes in a cascade---equally, the number of nodes in a causal tree;
\item $T$, the time span of a cascade: the length of the time interval between the first spike and the last;
\item $G$, the generation number, which is the number of levels of the tree---also, the length of the longest chain of successive excitations.
\end{itemize}
We are so equipped with the tools to analyze the statistics of cascades in the neuronal network: the function $H$ under considerations will be either one of $S,T,G$. As mentioned in Sect. \ref{ssec-invm}, we replace phase--space averages by Birkhoff sums. We consider time-segments of length $L = 4  \cdot 10^6$ iterations, following an initial transient (discarded) consisting of 6000 iterations. Occasionally we raise this number to account for the fact that long transient can be observed in extended systems \cite{politi1}. Using a parallel architecture, we accumulate the results of $32$ realizations of the initial conditions. We performed an extensive investigation, of which we report here a typical instance.

\begin{example}
$N=3000$ neurons, with  $n_{ex}=2400$, $m_{ex}=2$, $n_{in}=600$, $m_{in}=1$, $p_{ex}=4$, $p_{in}=2$, Physical parameters are: $\beta=0.133$,
$\alpha=3.6$, $\sigma_{\mbox{\tiny NINS}}=0.09$, $\sigma_{\mbox{\tiny ISN}}=0.103$, $\mu = 10^{-3}$, $\psi=3$, $\chi_{\mbox{\small exc}} = 0$, $\chi_{\mbox{\small inh}} = -1.1$.
\label{ex-one}
\end{example}

\begin{figure}
\includegraphics[width=.6\textwidth, angle=270]{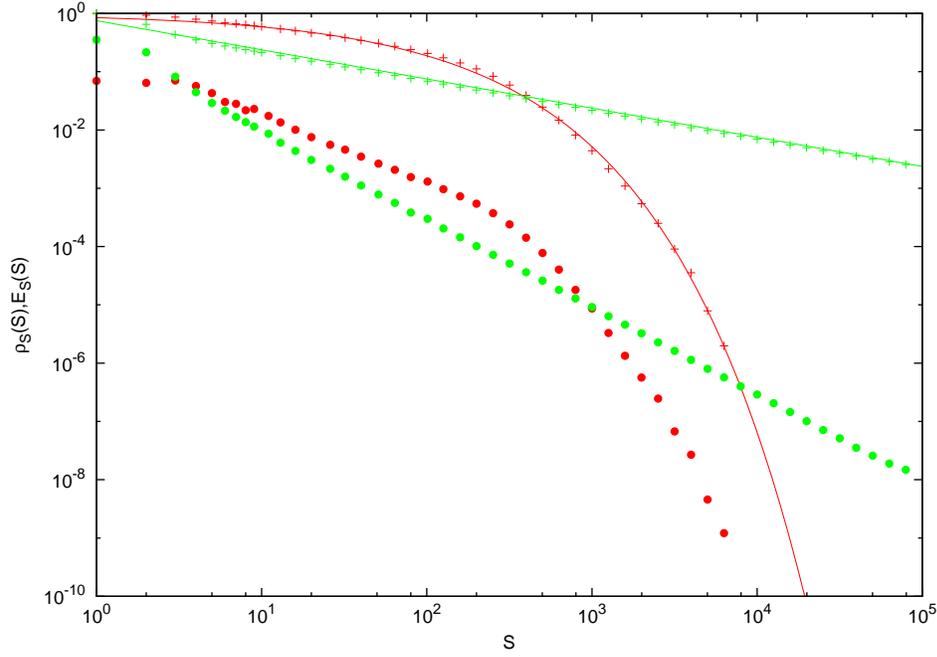}
\caption{Cascades statistics in the case of Example \ref{ex-one}. Fraction of cascades with $S$ spikes, $\rho_S(S)$, versus $S$, for $W=0.087$ (red bullets) and $W=0.09$ (green bullets) and the complementary distribution functions $E_S(S)$ (red and green crosses, respectively). The curves fitting these latter (continuous red and green) are described in the text. }
\label{fig-f162}
\end{figure}

Since this model neither includes pruning nor adaptation, to achieve criticality one needs to tune the parameters of the network. Following \cite{ruedi2}, we vary the coupling intensity $W$ in Eq. (\ref{eq-fir1}).
We first compute the  {\em average number of events per-unit time} $\tau^{-1}$, observing a rather sharp transition as we increase $W$:
when $W=0.084$ one has $\tau^{-1}=3.6  \cdot 10^{-2}$, which increases to $\tau^{-1}=5.68  \cdot 10^{-1}$ for $W=0.087$ and to $\tau^{-1}=5.65$ for $W=0.09$. At the largest coupling, roughly five neurons (out of the total 3000) spike on average at each single (discrete) time.
This transition is clearly reflected in the statistics: in Fig. \ref{fig-f162} we plot the fraction $\rho_S(S)$ of firing cascades with $S$ spikes and its (complementary) distribution function $E_S(S)$ defined in Eq. (\ref{eq-disth2}).
Two values of the coupling constant $W$ yield the data in Fig. \ref{fig-f162}: $W=0.087$  and $W=0.09$.
The former is visibly an under-threshold case, the latter displays a full-fledged cascade which extends over orders of magnitude. In the first case the distribution $E_S(S)$ is well fitted by a stretched exponential: $E_S(S)\simeq  \exp \{ -A S^B \}$ with $A \simeq -.1694$, $B \simeq .497$ (red curve). In the second, the power-laws $\rho_S(S) \simeq C' S^{-\beta_S}$ and $E_S(S) \simeq C S^{-\beta_S+1}$ are well satisfied, with exponent $\beta_S = 1.5$ ($C=0.75$, green line).

\begin{figure}
\includegraphics[width=.6\textwidth, angle=270]{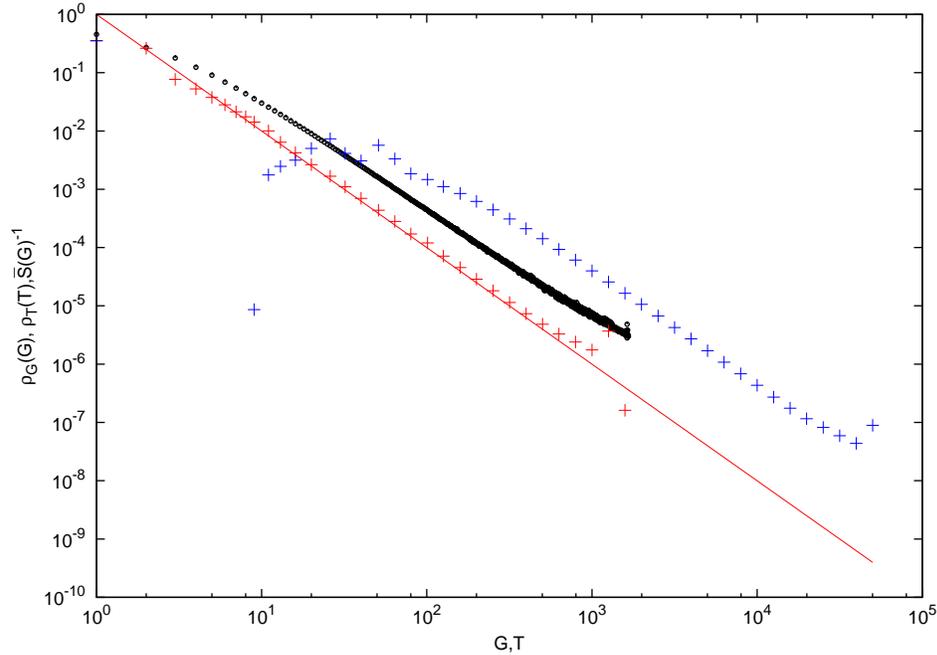}
\caption{Fraction $\rho_G(G)$ of cascades comprising $G$ generations (red crosses) and fraction $\rho_T(T)$ of cascades extending over $T$ units of time (blue crosses), for $W=0.09$. The red line is the power--law decay $\rho_G(G)= G^{-2}$. Also plotted is the inverse of the quantity $\bar{S}(G)$ versus $G$ (black bullets).}
\label{fig-66d1}
\end{figure}

In the case of large coupling, $W=0.09$, we plot in Fig. \ref{fig-66d1} the statistical distribution $\rho_G(G)$ of generation levels $G$ and of time span $T$, $\rho_T(T)$. For both of these quantities we observe a power--law decay with exponents $\beta_G$ and $\beta_T$ close to two. In the same figure we also plot the inverse, for graphical convenience, of the quantity $\bar{S}(G)$, the average number of spikes in a cascade with $G$ levels, which approximately follows a power--law with the same exponent: $\bar{S}(G) \sim G^\gamma$, $\gamma \simeq 2$.

The observed exponents $\beta_S = 3/2$ and $\beta_T=2$ are to a good approximation equal to those appearing in physiological experiments \cite{plenz03,mazzoni} and in various models of neuronal networks \cite{ott16,lucilla1}. To the contrary, these values differ from those obtained in \cite{ruedi2} for the system that we are investigating using the conventional definition of cascade.
It is also to be quoted the measurement of these exponents in high resolution experimental data \cite{beggs12}, still within the time--binning approach.

Critical scaling theory predicts \cite{sethna,beggs12} that $\beta_T - 1 = (\beta_S -1 ) {\gamma'}$, where $\gamma'$ is defined in terms of the scaling of the average of $S$ at fixed time length $T$, versus $T$. The exponents derived in \cite{ruedi2} verify this scaling relation. Since in our model the time span of a cascade $T$ and its generation level $G$ are proportional, we can identify the exponents $\gamma$ and $\gamma'$, providing the relation
\beq
 \rho_S(S) \sim S^{\beta_S}, \;\; \beta_S = \frac{\beta_G + \gamma - 1}{\gamma},
\nuq{eq-aves4}
From the data of Fig. \ref{fig-66d1} we see that approximately $\gamma=2$, so that the relation (\ref{eq-aves4}) for the triple $\beta_S = 3/2$, $\beta_G=\gamma=2$ is rather well satisfied by the numerical data.

\section{Statistics of extreme events in a finite time--interval}
\label{sec-evt}
The excitation of an intrinsically spiking neuron propagates sometimes to a large number of neurons and generates a cascade. In the previous section we have  studied the relative frequency of these cascades versus their intensity, in a theoretically infinite time--interval. It is now interesting to consider the probability that none of these cascades, within a certain {\em finite} observational interval, exceeds a certain intensity threshold---equivalently, the probability that the {\em largest} cascade in a time--interval is smaller than such threshold. This investigation belongs to the realm of extreme value theory and goes under the name of statistics of {\em block maxima}.

Let in fact $n$ be the time at which an excitatory ISN fires and generates a cascade (possibly, and frequently, limited to such single spike). To this value we associate an intensity, $I(n)$, which can be chosen to be the size of the cascade $S$, its level $G$, or its time span $T$. If more than a cascade is originated at time $n$, $I(n)$ is the maximum of the values corresponding to such cascades. To the contrary, if {no} cascade starts at time $n$ ({\em i.e.} no excitatory ISN spikes at time $n$) $I(n)$ is null.
Since the system is deterministic, $I(n)$ is actually also a function of the phase--space point $z$ at which the trajectory is located when one starts recording these intensities:
\beq
  I(z,n) = H(z_n) = H (\varphi^n (z))
\nuq{eq-estr1}
Following standard usage we then define the block maxima function ${H}_L(z)$, for various values of $L$, as the largest intensity of a cascade that originates in the time--window $n=0,\ldots,L-1$:

\begin{equation}
\begin{split}
{H}_L(z)&= \max\{I(z,0),I(z,1),\ldots,I(z,L-1))\}\\
 &= \max\{H(z),H(\varphi (z)),\ldots,H(\varphi^{L-1}(z))\}.
\end{split}
\end{equation}\label{eq-mn}

The cascade achieving maximum intensity is therefore appropriately called the {\em extreme event} in the time window under consideration; the wider is the time--window, the larger is presumably its intensity.
An \emph{Extreme Value Law} for ${H}_L(z)$ is an asymptotic form, for large $L$, of the statistical distribution of this variable. We now show that it applies to the present case.

Let us consider the distribution of the values of ${H}_L(z)$ with respect to an {invariant measure} $\mu$. As discussed before, we assume that the distribution function $F_{{H}_L} (h) = \mu (\{ z \in Z \mbox{ s.t. } H_L(z) \leq h \})$ can be obtained numerically by Birkhoff time averages over trajectories of the system.

Suppose that different cascades arise independently of each other, that is, correlation of the values of $H(\varphi^j(z))$ for different $j$ decays quickly when their time difference increases. 
Under this assumption the probability that ${H}_L(z)$ is less than a certain value $h$, which is clearly equal to the probability that the intensity $H(\varphi^n(z))$ of {\em all} cascades in the time interval $n \in [0,L-1]$ is less than $h$, becomes a {\em product} of the probabilities of these individual events. If we let $\lambda$ be the density of cascades per unit time (not to be confused with the density of spikes per unit time), we expect a number $\lambda L$ of cascades to originate in a time--window of length $L$, so that
\beq
F_{{H}_L}(h) \sim (F_H(h))^{\lambda L}.
\nuq{eq-scaling}
Next, taking logarithms and letting $h$ grow, so that $F_H(h)$ approaches one, we obtain
\beq
\begin{split}
\log (F_{{H}_L}(h)) &\sim \lambda L \log (F_H(h))\\
                             &\sim \lambda L (F_H(h)-1)\\
                             &= -\lambda L E_H(h).
\end{split}
\nuq{eq-eval1}
Clearly, this is a heuristic estimate, but it yields the same results predicted by rigorous theory, which requires appropriate conditions \cite{lucabook} to dispose of the correlations of the motion that might render the independence assumption incorrect. We can verify {\em a posteriori} the fact that these correlations fade in the limit of large $L$ by a direct evaluation of Eq. (\ref{eq-eval1}), when the asymptotic relation symbol is replaced by a much more stringent equality.

\begin{figure}
\includegraphics[width=.6\textwidth, angle=270]{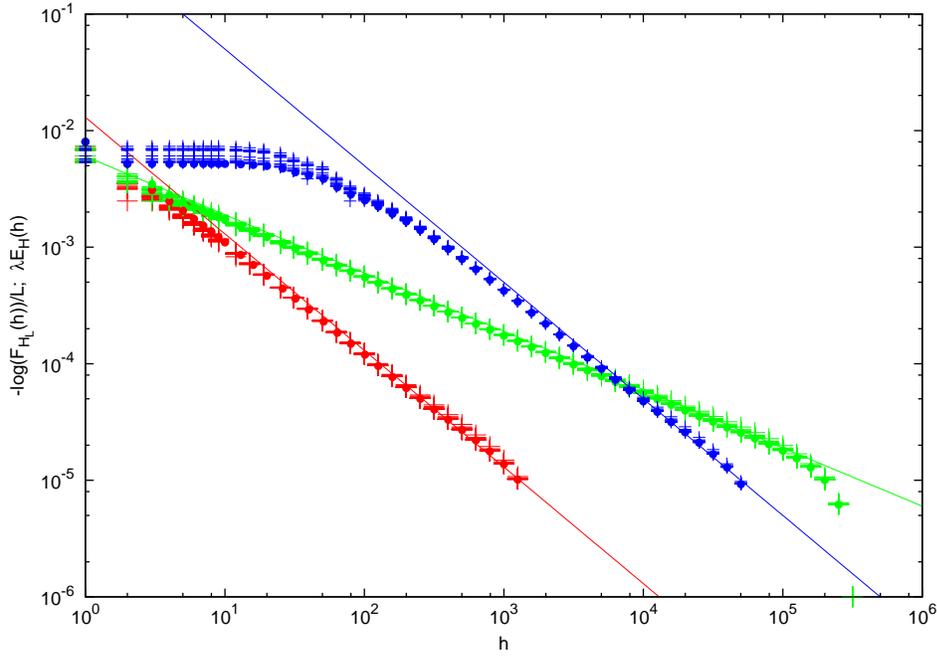}
\caption{Extreme value laws in the case of Example \ref{ex-one}, when $W=0.09$. Plotted are the values of  $-\log (F_{{H}_L}(h))/L$ (crosses) and $\lambda E_H(h)$ (bullets) versus $h$, when $H$ is $G$ (red), $S$ (green) and $T$ (blue). Values of $L$ range from $10$ to $10^4$.
Also displayed are the power--law decay laws: $0.013 \cdot h^{-1}$ (red), $0.006 \cdot h^{-1/2}$ (green) and $0.5 \cdot h^{-1}$ (blue). }
\label{fig-f162b}
\end{figure}

To do this, let us consider again the case of Example \ref{ex-one} when $W=0.09$. Dividing the total number of cascades by the total time span of the computation (averaged over $32$ different runs) yields the value $\lambda = 8 \cdot 10^{-3}$, which is much smaller than $\tau^{-1} = 5.65$ and reveals that spikes are organized in cascades of average size $(\lambda \tau)^{-1}$, that is, roughly $22$ spikes per cascade, in this case.
In Fig. \ref{fig-f162b} we plot the quantity $-\log (F_{{H}_L}(h))/L$ versus $h$, for various values of $L$, for the three functions $H=G,S,T$. In all three cases data points collapse on a curve, coinciding with $\lambda E_H(h)$, as predicted by Eq. (\ref{eq-eval1}). Observe that the value of $\lambda$ employed in the figure does not come from a fit of the data, but it is the one just obtained by counting the number of cascades. The extreme value law given by Eq. (\ref{eq-eval1}) appears therefore to apply, at least in this regime.

It is to be noted that in certain cases an additional positive factor $\theta$, smaller than one, appears in front of $\lambda E_H(h)$. This factor is called the extremal index: it implies a slower decay of $F_{{H}_L}(h)$, originating from the tendency of extreme events to appear in clusters. Rigorous theory has studied to a large detail this phenomenon in simple systems \cite{jorge}, which has also been detected in many dimensional systems \cite{nature}. The data of Fig. \ref{fig-f162b} indicate a unit value of this parameter.

\begin{figure}
\includegraphics[width=.6\textwidth, angle=270]{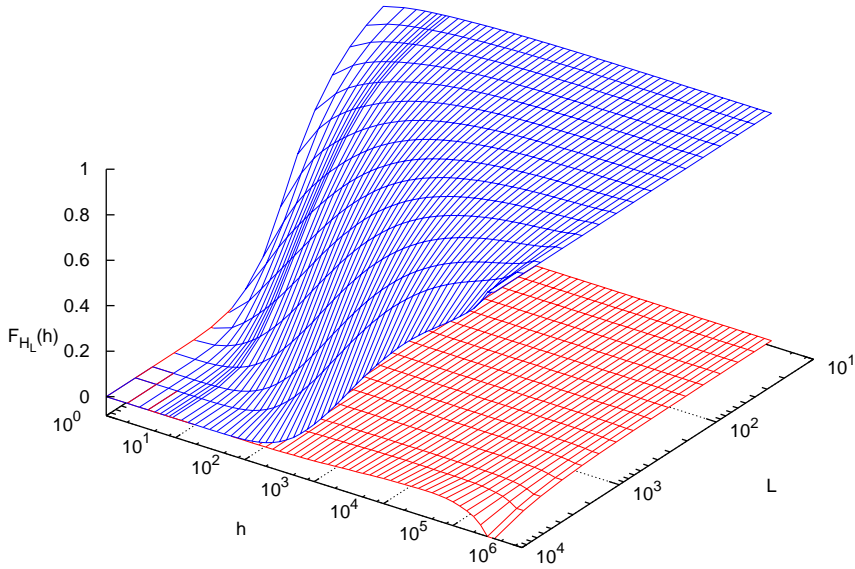}
\caption{The experimental function $F_{{H}_L}(h)$ for $H=S$ versus $L$ and $h$ in the case of Example \ref{ex-one} and W=0.09 (blue graph). Difference between the experimental function and the theoretical expression (\ref{eq-eval11}) (red graph). }
\label{fig-pro74b}
\end{figure}

In the critical regime the three quantities $H=G,S,T$ are characterized by a power--law decay: $ E_H(h) \simeq A_H h^{-\beta_H}$, where $A_H$ are suitable constants and the exponents $\beta_H$ have been discussed in the previous section. Inserting this information in Eq. (\ref{eq-eval1}) yields the relation
\beq
F_{{H}_L}(h)  \simeq \exp\{ - \lambda L A_H h^{-\beta_H} \}.
\nuq{eq-eval11}
In Fig. \ref{fig-pro74b} we plot the numerically determined function $F_{{H}_L}(h)$ when $H=S$ for values of $L$ and $h$ ranging over orders of magnitude and their difference with the analytical relation (\ref{eq-eval11}). This latter is remarkably small in all the range of data considered.

\begin{figure}
\includegraphics[width=.6\textwidth, angle=270]{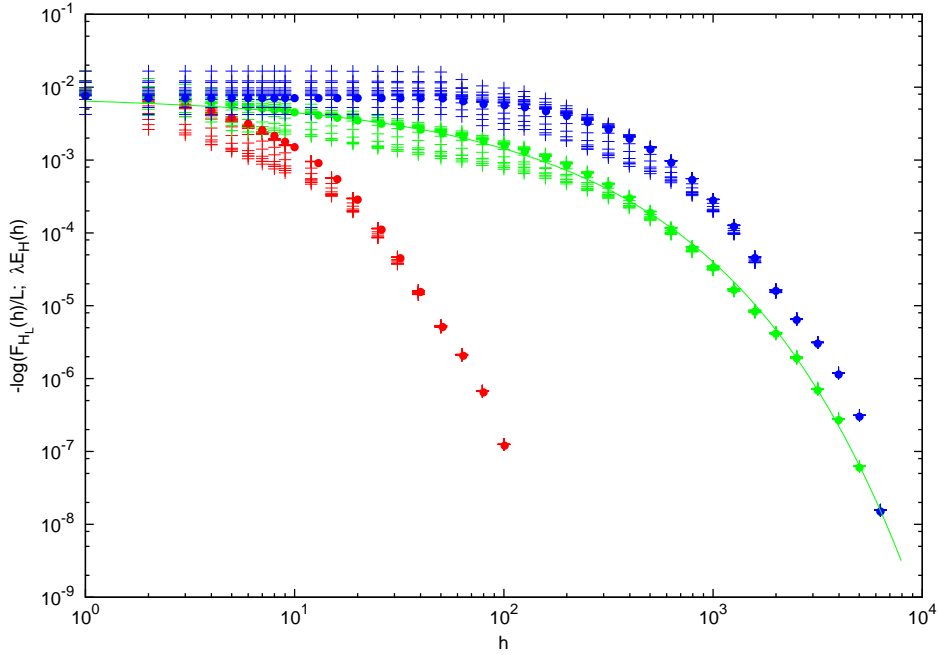}
\caption{Extreme value laws in the case of Example \ref{ex-one}, when $W=0.087$. The same data as in Fig. \ref{fig-f162b} are plotted: the quantities  $-\log (F_{{H}_L}(h))/L$ (crosses) and $\lambda E_H(h)$ (bullets) versus $h$, when $H$ is $G$ (red), $S$ (green) and $T$ (blue). Values of $L$ range from $10$ to $10^4$. Asymptotic coalescence of the data is observed. The green curve is the stretched exponential
$E_S(S)\simeq  \exp \{ -A S^B \}$ with $A \simeq -0.1694$, $B \simeq 0.497$ already
plotted in Fig. \ref{fig-f162} and described in Sect. \ref{sec-quantities}.}
\label{fig-run65a2}
\end{figure}

\begin{figure}
  \begin{center}
    \begin{tabular}{cc} \resizebox{80mm}{!}{\includegraphics[height=55mm, angle=-90]{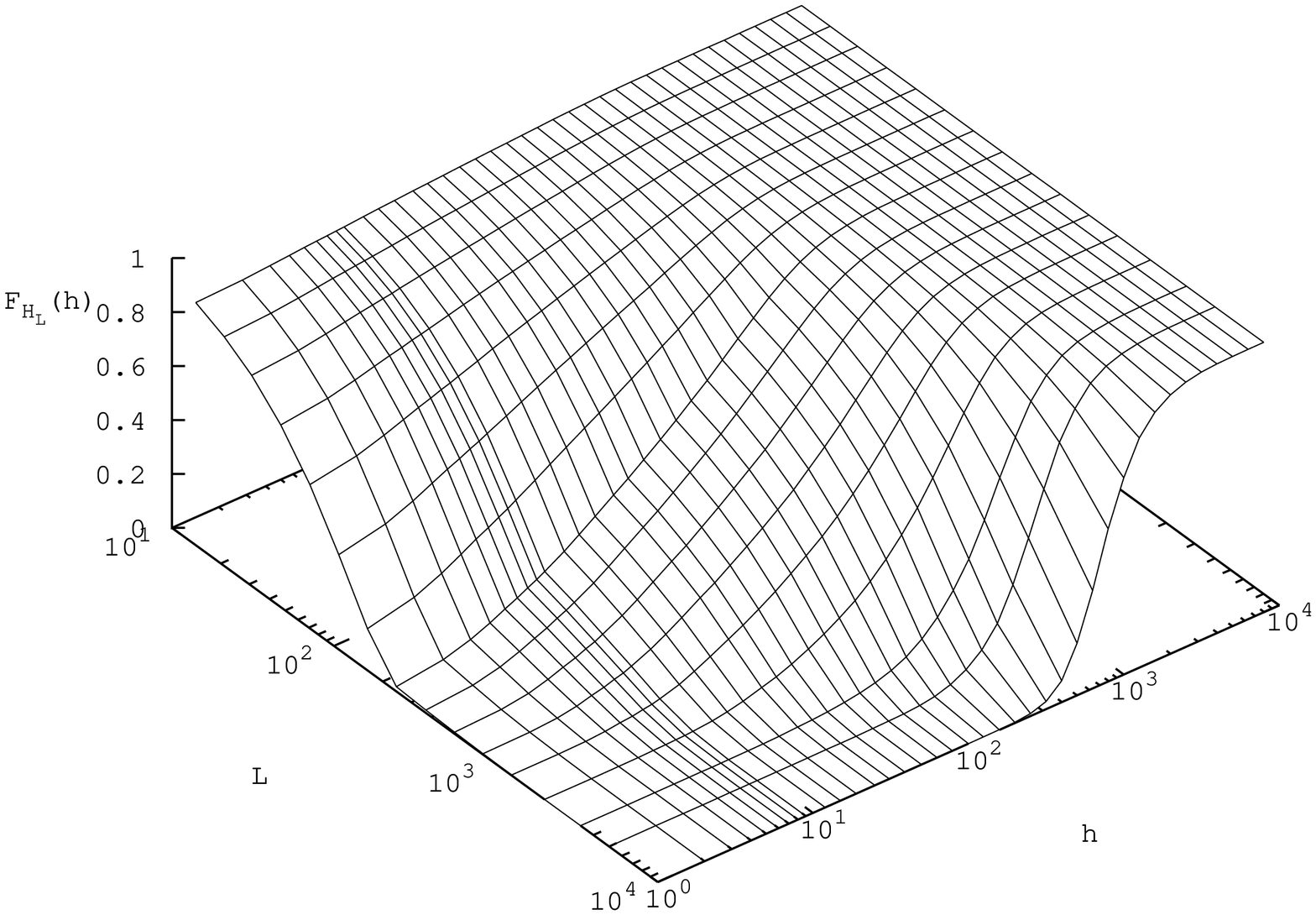}}&
      \resizebox{80mm}{!}{\includegraphics[height=55mm, angle=-90]{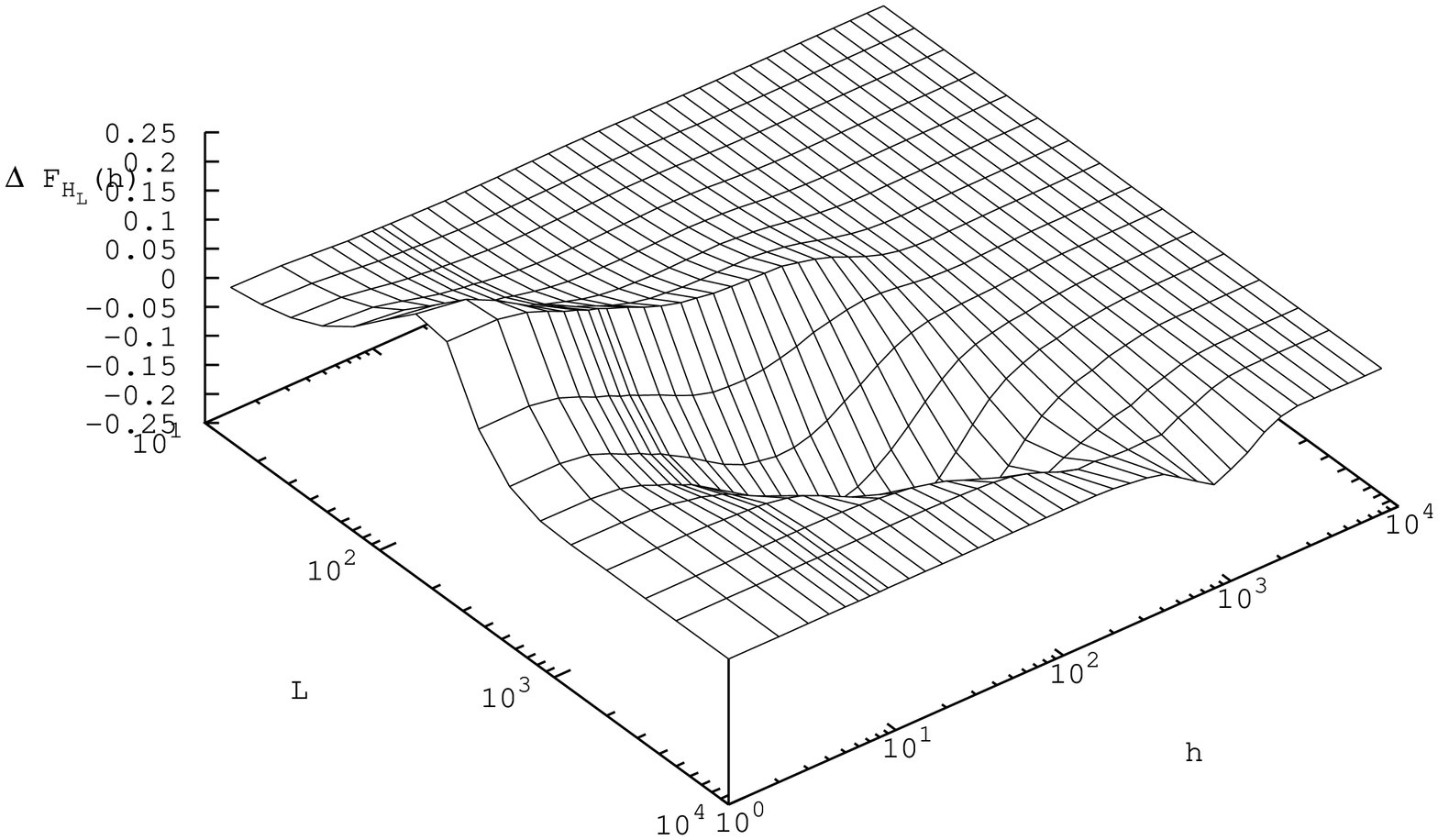}}
      \\
    \end{tabular}
  \caption{The experimental function $F_{{H}_L}(h)$ for $H=S$ versus $L$ and $h$ in the case of Example \ref{ex-one} and W=0.087 (left graph). Difference between the experimental function and the theoretical expression (\ref{eq-eval12}) (right graph). }
\label{fig-387-r65c}
  \end{center}
\end{figure}

We can perform the same analysis in the subcritical case, Example \ref{ex-one} and $W=0.087$. In Fig. \ref{fig-run65a2}, which is the analogue of Fig. \ref{fig-f162b} for this case, we plot again $-\log (F_{{H}_L}(h))/L$ and $\lambda E_H(h)$, for $H=G,S,T$. Collapse of the data points is again observed, but for larger values of $h$ than in the previous case. The asymptotic law for the distribution of the variable $H_L(h)$ becomes here
\beq
F_{{H}_L}(h)  \simeq \exp\{ - \lambda L \exp\{ A_H h^{B_H} \} \}.
\nuq{eq-eval12}
The experimental function $F_{{H}_L}(h)$ for $H=S$  is plotted versus $L$ and $h$ in the left frame of Fig. \ref{fig-387-r65c}. 
Difference with the theoretical expression (\ref{eq-eval12}) is plotted in the right frame. As expected from Fig. \ref{fig-run65a2}, convergence improves as $L$ and $h$ increase.

Within the framework of Extreme Value Theory, the two cases just discussed differ because they imply different {\em normalizing} sequences $h_L(t)$, which we now define, while providing the same limiting distribution $F(t)=e^{-t}$. In fact, observe that Eqs. (\ref{eq-scaling},\ref{eq-eval1}) imply that
\beq
F_{{H}_L}(h) \sim \exp\{ - \lambda L E_H(h)\}.
\nuq{eq-scaling2}
Let $t$ be a positive, real variable. The sequence of levels $h_L(t)$, $L=1,2,\ldots$, parameterized by $t$, is defined so to satisfy the limit behavior
\begin{equation}
\label{eq-un}
  \lambda L E_{H}(h_L(t)) \rightarrow t,\;\mbox{ as } L \rightarrow \infty.
\end{equation}
As a consequence of the above,

\begin{equation}\label{eq-evl}
\mu (\{ z \in Z  \mbox{ s.t. }  {H}_L(z) \leq h_L(t) \}) =F_{{H}_L}(h_L(t))
  \rightarrow F(t) =e^{-t}, \text{as } L\to\infty.
\end{equation}

The two previous equations are the core results of EVT, when added to the fact that {\em three} different limiting distributions $F(t)$ can be achieved after further rescaling, as we do below in the critical case.
They show that $h_L(t)$ is a quantile of the asymptotic distribution of the statistical variable $H_L(z)$, whose analytical form involves the ratio $t/L$ and the inverse of the complementary distribution of the function $H$:
 \begin{equation}
\label{eq-ev4}
  h_L(t) \sim E_H^{-1}( \frac{t}{\lambda L}).
\end{equation}
In the subcritical case these quantiles grow as the power of a logarithm:
 \begin{equation}
\label{eq-ev41}
  h_L(t) \sim \left( \log ( \frac{\lambda L}{t}) \right)^\frac{1}{B_H}.
\end{equation}
At criticality a power--law behavior is present:
 \begin{equation}
\label{eq-ev42}
  h_L(t) \sim ( \frac{\lambda L}{t}) ^\frac{1}{\beta_H}
\end{equation}
and any $q$ moment of the statistical distribution of $H_L(z)$ scales as $(\frac{\lambda L}{t}) ^\frac{q}{\beta_H}$.
In this latter case, the previous relations can be put in one of the three standard forms \cite{gnede}. In fact, simple algebra permits to rewrite
\begin{equation}\label{eq-ev8}
\mu (\{ z \in Z  \mbox{ s.t. }  (\lambda L)^{-1/\beta_H} {H}_L(z) \leq u \})
  \rightarrow e^{-u^{\beta_H}},
\end{equation}
as $L\to\infty$, which is the expression of a Fr\'echet distribution.

\section{Conclusions}
\label{sec-conc}
We have described in this paper an extension of the dynamical theory of extreme events to time--evolving phenomena in complex, multi--dimensional systems. We have applied it with success to a model of neuronal network, that was studied in full detail. This analysis attains wide generality because this example is characterized by features that are frequently encountered: it is composed of many almost identical subsystems, linked in a sparse, directed network; each individual system integrates the input from the network and in turn returns a signal to it; signal propagation can develop a full chain of excitations. This has lead us to introduce a novel approach to define neuronal avalanches, whose statistical characteristics were studied within the framework of extreme value theory. The same definition of avalanche and a similar analysis can be obviously applied to other systems in the class just described.

The presented results are in line with previous theoretical models and experimental findings on neuronal networks. Nonetheless, since the focus of this work is on dynamical theory, we did not investigate various phenomena of physiological relevance, such as the balance between excitatory and inhibitory neurons (that are nonetheless present in our model) and the r\^ole of pruning and adaptation to sustain criticality \cite{lucilla2}. Also, we did not perform rescaling of the shape of cascades at criticality, to reveal a single curve, as in crackling noise, although it is highly likely that this will happen, as shown in \cite{ruedi2}. A further topic that could be addressed is the computation of the branching ratio in the chain of excitations, which is clearly possible with the aid of our definition of cascade. And finally, we did not compare among them different network topologies.

While the set-up that we propose to extend the dynamical extreme value theory to complex phenomena, evolving in time, is fully rigorous, our results are formal, even if confirmed by numerical experiments. This poses a new challenge for rigorous mathematical analysis, while at the same time it offers a paradigm for experimental studies in various physical systems.

{\em \bf Acknowledgements} Numerical computations for this paper were performed on the Zefiro cluster at the INFN computer center in Pisa. Enrico Mazzoni is warmly thanked for assistance.

\section{Appendix: stability of the fixed point of Rulkov map }

It is easy to derive the fixed point of Rulkov map and its stability. Consider again an isolated neuron, and assume that we are in the first alternative of Eq. (\ref{eq-map1b}), which yields the map
\beq
 \left \{
\begin{array}{lcl}
    x & \rightarrow & \frac{\alpha}{1 -x} + y \\
          y & \rightarrow &  y - \mu (x+1) + \mu \sigma .
\end{array}
\right .
\nuq{eq-map1b2}
From this, the unique fixed point $(\bar x , \bar y)$ easily follows:
\beq
 \left \{
\begin{array}{lcl}
    \bar x  & = & \sigma - 1 \\
     \bar  y & = & \sigma - 1 - \frac{\alpha}{2 - \sigma}
\end{array}
\right .
\nuq{eq-map1b3}
This result is consistent with Eq. (\ref{eq-map1b}) provided that $\bar x < 0$, that is, $\sigma < 1$.

\begin{figure}
\centerline{\includegraphics[height=10cm, angle=270]{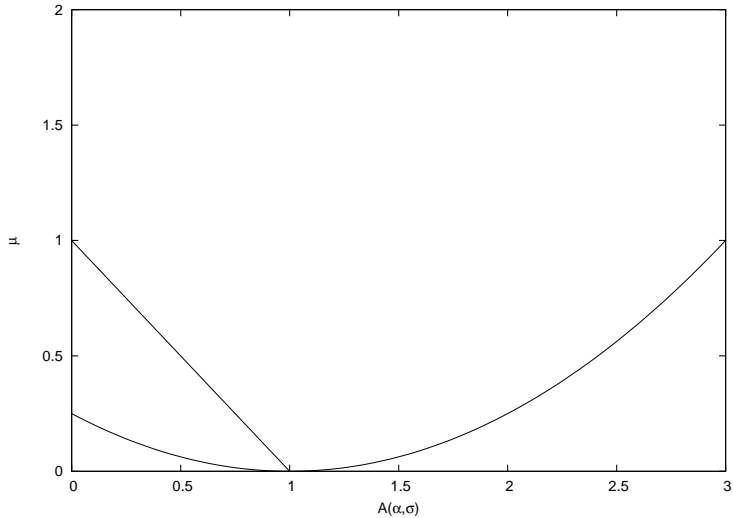}}
\caption{Stability regions of the Rulkov map (\ref{eq-map1b3}) in parameter space. Moving from the bottom left corner $(0,0)$ to the bottom right $(3,0)$ one encounters regions I to IV in order. See text for further details.}
\label{fig-regio}
\end{figure}

The differential of the transformation, {\em i.e.} the linear map from the tangent space at the fixed point to itself is
\beq
 J = \left [
\begin{array}{lcl}
   \frac{\alpha}{(2 - \sigma)^2}    &  &  1 \\
     -\mu &   & 1
\end{array}
\right ]
\nuq{eq-map1b4}
Putting for simplicity $A(\alpha,\sigma) = \frac{\alpha}{(2 - \sigma)^2}$ the linear map $J$ has real eigenvalues provided that $A(1-(\alpha,\sigma))^2 \geq 4 \mu$. In the opposite case, complex conjugate pairs appear. In this case, the square modulus of the eigenvalues equals the determinant of $J$ and is
\[
\mbox{Det }(J) = A(\alpha,\sigma) + \mu.
\]
This situation is pictured in Fig. \ref{fig-regio}: $ A(\alpha,\sigma)$ and $\mu$ are always positive and are taken as coordinates in the $(A,\mu)$ plane. When the corresponding point lies below the parabola $(1-A(\alpha,\sigma))^2 = 4 \mu$,  $J$ has two real eigenvalues. If moreover $A<1$ (region I) their absolute value is less than one and the (hyperbolic) fixed point is attractive, while it is repulsive when $A>1$ (region IV).
In the region above the parabola and below the line $A(\alpha,\sigma) + \mu = 1$ (region II) $J$ has a pair of complex conjugate eigenvalues of modulus less than one, so that the fixed point is still attractive, and trajectories spiral down to it. When finally $ A(\alpha,\sigma)$ is above both the parabola and the straight line (region III) two eigenvalues of modulus larger than one appear: the fixed point is unstable, and trajectories spiral out of it.

\end{document}